\documentclass[aps,pra,nofootinbib,twocolumn,superscriptaddress]{revtex4}
\usepackage{amsmath, amsthm, amssymb,graphicx,color,bbm,hyperref}
\begin{document}

\newcommand{\rrangle}{\rangle\!\rangle} \newcommand{\llangle}{\langle\!\langle}

\newcommand{\logL}{\log\mathcal{L}}

\newcommand{\expec}[1]{\ensuremath{\left\langle#1\right\rangle}}

\newcommand{\cM}{\mathcal{M}}
\newcommand{\cG}{\mathcal{G}}
\newcommand{\cD}{\mathcal{D}}
\newcommand{\cE}{\mathcal{E}}
\newcommand{\cL}{\mathcal{L}}
\newcommand{\cU}{\mathcal{U}}
\newcommand{\cH}{\mathcal{H}}

\newcommand{\reals}{\mathbb{R}}

\newcommand{\diff}{\mathrm{d}\!}
\newcommand{\pdiff}[2]{\frac{\partial #1}{\partial #2}}
\newcommand{\pdiffsq}[3]{\frac{\partial^2 #1}{\partial #2\partial #3}}
\newcommand{\expect}[1]{\ensuremath{\left\langle#1\right\rangle}}

\newcommand{\qhat}{\hat{q}}
\newcommand{\rhohat}{\hat{\rho}}
\newcommand{\rhoMLE}{\hat{\rho}_{\mathrm{MLE}}}
\newcommand{\rhotrue}{\rho_{\mathrm{true}}}
\newcommand{\Rhat}{\hat{\mathcal{R}}}
\newcommand{\Vbar}{\overline{V}}

\newcommand{\ket}[1]{\ensuremath{\left|#1\right\rangle}}
\newcommand{\bra}[1]{\ensuremath{\left\langle#1\right|}}
\newcommand{\braket}[2]{\ensuremath{\left\langle#1|#2\right\rangle}}

\newcommand{\ketbra}[2]{\ket{#1}\!\!\bra{#2}}
\newcommand{\braopket}[3]{\ensuremath{\bra{#1}#2\ket{#3}}}
\newcommand{\proj}[1]{\ketbra{#1}{#1}}

\newcommand{\sket}[1]{\ensuremath{\left|#1\right\rrangle}}
\newcommand{\sbra}[1]{\ensuremath{\left\llangle#1\right|}}
\newcommand{\sbraket}[2]{\ensuremath{\left\llangle#1|#2\right\rrangle}}
\newcommand{\sketbra}[2]{\sket{#1}\!\!\sbra{#2}}
\newcommand{\sbraopket}[3]{\ensuremath{\sbra{#1}#2\sket{#3}}}
\newcommand{\sproj}[1]{\sketbra{#1}{#1}}
\newcommand{\rbk}[1]{\textcolor{blue}{#1}}
\newcommand{\noemph}[1]{#1}
\def\Id{1\!\mathrm{l}}
\newcommand{\Tr}{\mathrm{Tr}}
\newcommand{\Nparams}{N_{\mathrm{params}}}

\title{Demonstration of qubit operations below a rigorous fault tolerance threshold with gate set tomography}
\begin{abstract}
Quantum information processors promise fast algorithms for problems inaccessible to classical computers.  But since qubits are noisy and error-prone, they will depend on fault-tolerant quantum error correction (FTQEC) to compute reliably.  Quantum error correction can protect against general noise if -- and only if -- the error in each physical qubit operation is smaller than a certain threshold.  The threshold for general errors is quantified by their diamond norm.  Until now, qubits have been assessed primarily by randomized benchmarking, which reports a different ``error rate'' that is not sensitive to all errors, and cannot be compared directly to diamond norm thresholds.  Here we use gate set tomography (GST) to completely characterize operations on a trapped-Yb$^+$-ion qubit and demonstrate with very high ($>95\%$) confidence that they satisfy a rigorous threshold for FTQEC (diamond norm $\leq6.7\times10^{-4}$).
\end{abstract}

\author{Robin Blume-Kohout}
\affiliation{Center for Computing Research, Sandia National Laboratories, Albuquerque, NM 87185, USA}
\author{John King Gamble}
\affiliation{Center for Computing Research, Sandia National Laboratories, Albuquerque, NM 87185, USA}
\author{Erik Nielsen}
\affiliation{Sandia National Laboratories, Albuquerque, NM 87185, USA}
\author{Kenneth Rudinger}
\affiliation{Center for Computing Research, Sandia National Laboratories, Albuquerque, NM 87185, USA}
\author{Jonathan Mizrahi}
\altaffiliation[Current Address:]{ Joint Quantum Institute, University of Maryland Department of Physics and National Institute of Standards and Technology, College Park, MD 20742}
\affiliation{Sandia National Laboratories, Albuquerque, NM 87185, USA}
\author{Kevin Fortier}
\affiliation{Sandia National Laboratories, Albuquerque, NM 87185, USA}
\author{Peter Maunz}
\affiliation{Sandia National Laboratories, Albuquerque, NM 87185, USA}

\date{\today}
\maketitle

\begin{figure*}[tb]
\includegraphics[width= 1.0 \linewidth]{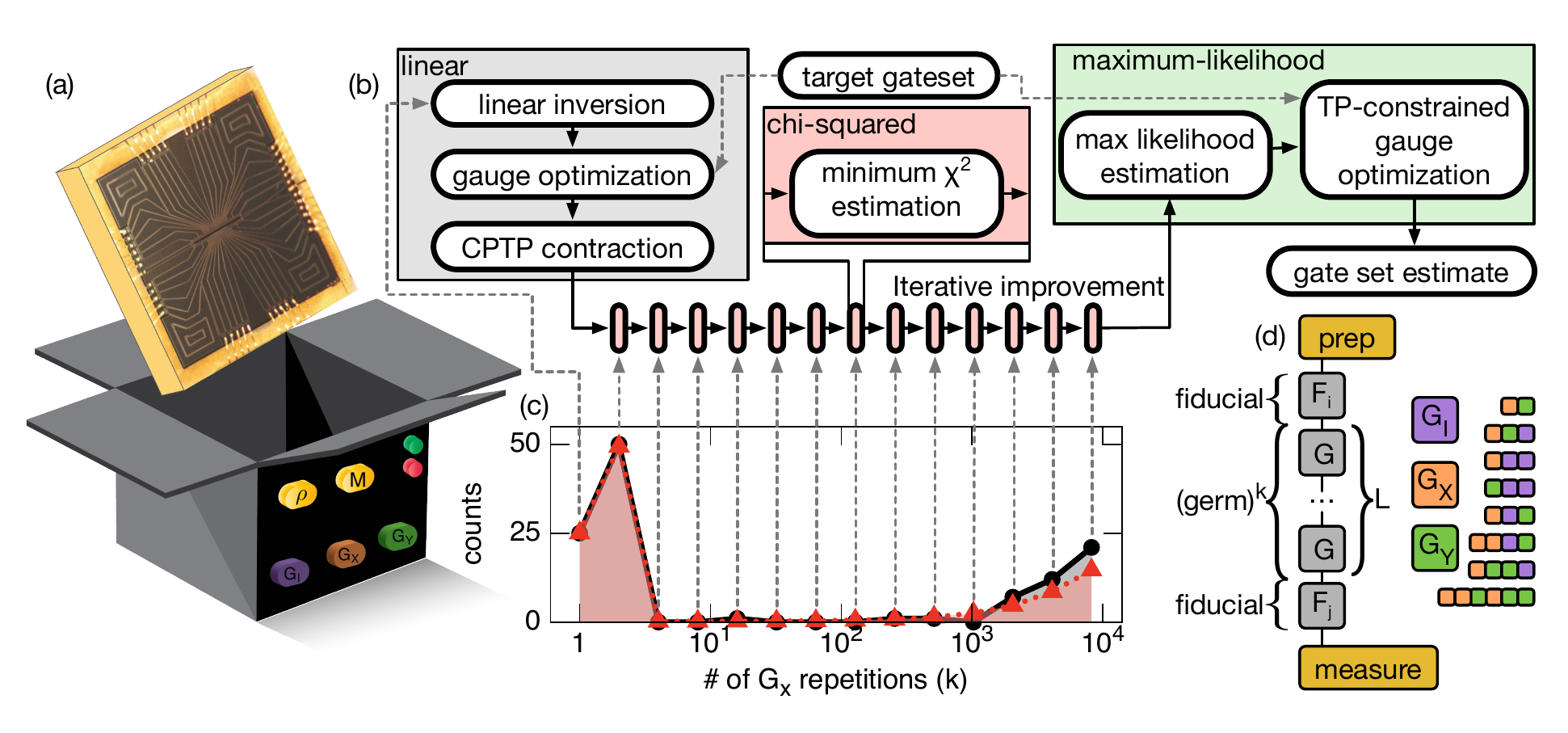}
\caption{\label{fig:schematic}
\textbf{Gate set tomography is a robust procedure to characterize as-built qubits.} 
\textbf{(a)}: Gate set tomography (GST) models the qubit (a single Yb$^+$ ion localized in a linear surface electrode ion trap)
as a ``black box" with a classical interface, and is agnostic to its physical details.
\textbf{(b)}: Flow chart of GST analysis. Its core is an iterative $\chi^2$ minimization, over 
data from increasingly long gate sequences, seeded with a linear inversion estimate.  The 
final step, likelihood maximization, produces an accurate and unbiased estimate of the gate set.
\textbf{(c)}: A subset of the nearly 5000 data points taken:  observed ``bright'' counts (black) for sequences of the form $G_x^k$ and the GST estimate's prediction (red; see Fig.~\ref{fig:GSTResults}).  Deviations from ideal gates appear only at $k>1000$. 
\textbf{(d)}: GST achieves high precision from periodic sequences based on short ``germs''.
Here, the 11 germs used for this experiment are shown (colored boxes), as is the ``fiducial 
sandwich'' form of a general GST sequence.}\end{figure*}

\section{Introduction}
The field of quantum information processing has seen great growth over the past thirty years, driven by exciting quantum algorithms inaccessible to classical computers.  
Small quantum information processors have been demonstrated experimentally using
superconducting circuits \cite{nakamura:1999,wallraff:2004,barends:2014},
electrons in semiconductors \cite{petta:2005,shulman:2012,veldhorst:2015},
trapped atoms and ions \cite{leibfried:2003,barrett:2004,reiserer:2014},
and photons \cite{obrien:2003,peruzzo:2010,politi:2009}.
Trapped ions are among the most reliable qubits available today; up to 14 qubits have been addressed in a single trap \cite{MonzPRL11}, a 5-qubit quantum information processor has been realized \cite{Debnath2016}, and single-qubit gates have demonstrated randomized benchmarking infidelities as low as $10^{-6}$ \cite{harty:2014,BrownPRA11,Mount2015}.

Unlike classical bits, qubits are intrinsically noisy and error-prone, and will require active, fault-tolerant quantum error correction (FTQEC \cite{QEC2013}) to operate reliably. 
To function, FTQEC requires physical qubit operations to be high quality, with errors below a specific \noemph{threshold}.
Fault tolerance (FT) thresholds for quantum computing have been proven against various noise models, and generally require per-gate failure rates between $10^{-6}$ and $10^{-2}$ \cite{knill:2005,Aliferis:2006,aliferis:2007,Aliferis:2009}.  
However, the particular metric of ``error rate" depends on the noise model.  Against realistic general errors, including small unitary errors, thresholds are stated in terms of the gates' \noemph{diamond norm} error, $||G_i - G_i^{(ideal)}||_\diamond$ \cite{AliferisNJP09,Aliferis:2006,Aharonov:2008}.

Randomized benchmarking (RB) \cite{EmersonScience2007,Knill2008}, the most commonly used method for qubit characterization, measures a single error rate ($\epsilon_{\mathrm{RB}}$) that closely approximates the gates' average process infidelity.  
Because RB is relatively insensitive to unitary errors \cite{SandersNJP16} that dominate diamond norm error \cite{Kueng15} and have unpredictable consequences for FTQEC \cite{Aliferis:2006}, it cannot efficiently measure diamond norm error to high precision.  This makes it nearly impossible to demonstrate suitability for fault tolerance using RB alone, unless errors are \noemph{assumed} to be strictly incoherent.
There are variants of RB that characterize and report additional parameters, but none of them are well suited for diamond norm characterization or comparison to fault tolerance thresholds \cite{MagesanPRL12,kimmel13,WallmanNJP15}.

We use a characterization method called gate set tomography (GST) \cite{GST2013,Greenbaum15,GST2015} to systematically debug and improve a 1-qubit trapped-Yb$^+$-ion quantum information processor, and -- finally -- to demonstrate with very high confidence that all three of its quantum logic operations surpass a proven threshold for FTQEC.
GST provides a full and extremely accurate tomographic description of every gate, complete with statistical confidence bounds.  We use this information to iteratively improve our single qubit operations and to place tight bounds on the diamond norm error of the final gates, producing the first single-qubit gates whose errors are demonstrably below a rigorous threshold for fault-tolerant error correction.

This is not a demonstration of FTQEC, which requires not just single-qubit gates, but also high-fidelity two-qubit gates, repeatable measurements, and (of course) more qubits.  However, the GST methods that we use here to demonstrate 1-qubit gate errors below the threshold do generalize to 2-qubit gates, to the characterization of repeatable measurements, and to important properties of multiqubit systems such as crosstalk.  So, while pushing single-qubit gate errors below the threshold is only one step toward achieving FTQEC, it is an important one.

\section{Results}
\label{sec:results}

\subsection{Gate set tomography}

Our goals are (1) to implement quantum operations satisfying a FTQEC threshold, and (2) to ``prove'' -- i.e., demonstrate conclusively -- that we have done so.  Genuine proofs are the domain of mathematics.  In experimental science, the highest achievable standard is to provide experimental data (or summary statistics) that: (1) are consistent with the desired outcome; and (2) are inconsistent with any other plausible theory, and thus rule out all alternatives to some high level of confidence.  Our intent is not to provide an \noemph{exclusive} protocol for such demonstrations, but rather to establish that gate set tomography is \noemph{sufficient} to do so.

While RB and quantum process tomography \cite{ChuangJMO97,WeinsteinJCP04,ObrienPRL04} can be used in this fashion, they each face nontrivial obstacles.  RB's insensitivity to unitary errors makes it a poor tool for bounding worst-case error rates (diamond norms).  In process tomography \cite{ShulmanScience12,MerkelPRA13}, small calibration errors in the gates used to implement different measurements propagate to the final results, invalidating them.  Gate set tomography (GST) is a self-calibrating tomography protocol that solves both of these problems.  GST protocols based on short quantum circuits were developed at IBM \cite{MerkelPRA13} and Sandia \cite{GST2013}.  The long-sequence GST protocol demonstrated here is orders of magnitude more precise.

GST relies on two assumptions: (a) the system being characterized is a qubit with a 2-dimensional Hilbert space; (b) each gate operation is stationary and Markovian.  It treats the qubit as a black box with operation buttons (one for initialization, one for measurement, and the rest for gate operations) as shown in Fig.~\ref{fig:schematic}a, and self-consistently determines all operations up to a choice of basis (a gauge; see Methods).  It can also detect and quantify violations of these assumptions (see next subsection).

In GST, the real (noisy) gates are modeled as trace-preserving linear maps on density matrices (TP maps).  Such maps must be \noemph{completely positive} to be physical, and thus are usually referred to as ``CPTP maps''; for technical reasons, we do not always impose the CP constraint in GST, but otherwise these maps are functionally the same as CPTP maps.

The qubit's quantum state $\rho$ is a 4-element vector $\sket{\rho}$ in the vector space of $2\times2$ Hermitian matrices (Hilbert-Schmidt space) \cite{GST2013}, and each gate is a $4\times 4$ matrix $G$ that acts on $\sket{\rho}$ by left multiplication (i.e., $\sket{\rho} \to \sket{\rho_t} = G\sket{\rho}$).  Measurement is represented by a 2-outcome positive operator-valued measure (POVM) $\{E,\Id-E\}$.  Our target state preparation and measurement (SPAM) are $\rho^{(ideal)} = \proj{0}$ and $E^{(ideal)} = \proj{1}$.

Data for GST come from gate sequences (quantum circuits), each comprising: (1) initialization, (2) a series of gates, and (3) measurement.
Each sequence is repeated $N$ times, and the frequency of 0/1 counts is recorded.  In the experiments reported here, we implemented and used the set of gates $\{G_I,G_X,G_Y\}$, but GST can analyze any gate set rich enough to prepare an informationally complete set of probe states and measurements.

GST analysis proceeds as shown in Fig.~\ref{fig:schematic}b.  First, a specific set of short sequences is analyzed by linear inversion (see Methods) to get a rough estimate of the gates and SPAM operations.  This estimate has an unavoidable gauge freedom; every observable probability is invariant under
\begin{align}
G_k \to M G_k M^{-1}\\
\sket{\rho} \to M \sket{\rho},
\sbra{E} \to \sbra{E}M^{-1},
\end{align}
for any invertible matrix $M$.  We choose a gauge that makes the estimated gates as similar to the target gates as possible (see Methods).  If the rough estimate is not already completely positive, we truncate each gate to the nearest CP map, to ensure physically valid probabilities in the next step.  
Next, using the rough estimate as a starting point, we iteratively add more data.  In the $m^{\text{th}}$ iteration, we add data from gate sequences of length $2^{(m-1)}$ into the pool, then numerically adjust the estimate to minimize the $\chi^2$ divergence between the observed frequencies and estimated probabilities.  This ``min-$\chi^2$'' estimate is then used as the seed for a numerical maximization of the likelihood function $\mathcal{L}(\hat{G}) = \mathrm{Pr}(\mathrm{data}|\hat{G})$.  Finally, we perform another gauge optimization to maximize similarity to targets.

The GST gate sequences (see Fig.~1d) are chosen to (collectively) amplify every physical parameter in the gate set.  Short sequences called \noemph{germs} are repeated many times, and these ``germ power'' sequences are pre- and post-fixed by each of six \noemph{fiducial sequences}. 
In this work, we use six fiducial sequences, $\{ \emptyset,G_X, G_Y, G_X G_X G_X, G_Y G_Y G_Y, G_X G_X \}$, where $\emptyset$ denotes the null sequence,
and $G_X$ ($G_y$) are noisy $\pi/2$ rotations about $x$ ($y$).  These fiducials map $\rho$ and $E$ to (approximately) the six Pauli eigenstates, defining an informationally complete experimental reference frame.
For further details on sequence design and a complete list of all experiments performed, see Methods.

\subsection{Experiment}

Our qubit is a single $^{171}$Yb$^+$ ion in a state-of-the-art linear surface ion trap (Fig.~\ref{fig:schematic}a). 
Ions are trapped by photoionizing neutral ytterbium vapor that reaches the trapping volume through a slot from the back of the surface trap chip.  The qubit is encoded in the hyperfine clock states of the  $^{2}S_{1/2}$ ground state of $^{171}$Yb$^+$: $\ket{0} = \ket{F=0, m_F=0}$,  $\ket{1}=\ket{ F=1, m_F=0}$ .  
Standard laser cooling techniques are applied to Doppler cool the ion and prepare it in the $|0 \rangle$ state \cite{olmschenk_manipulation_2007}.  Standard fluorescence state detection \cite{olmschenk_manipulation_2007} is used to measure in the $\{\ket{0},\ket{1}\}$ basis. 

Three logic gates -- $G_I$ (the idle or identity gate), $G_X$ (a $\pi/2$ $X$ rotation), and $G_Y$ (a $\pi/2$ $Y$ rotation) -- are realized by using a microwave horn to apply pulses near-resonant with the $12.6428\operatorname{MHz}$ separation of the qubit levels.  Broadband composite pulses (BB1)~\cite{wimperis_broadband_1994, merrill_progress_2012} are employed to minimize sensitivity to amplitude fluctuations in the microwave signal.  

\begin{figure}[tb]
\includegraphics[width= 1.0 \linewidth]{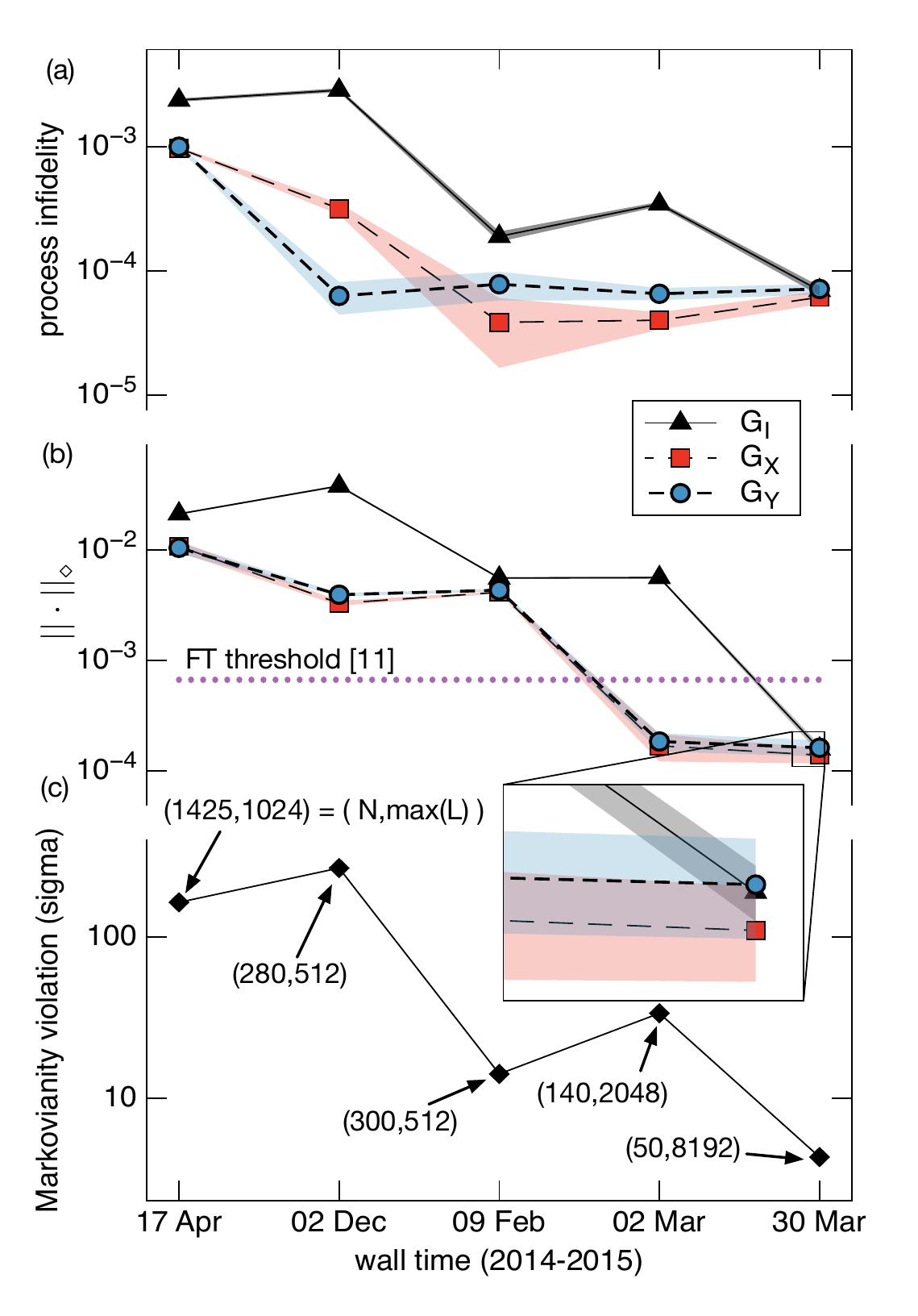}
\caption{\label{fig:resultsOverTime}
\textbf{Progressive improvement of quantum operations.}
Over the course of $\sim$1 year, we used GST to improve our qubit, ending with clearly sub-threshold error rates.  All metrics are computed using GST estimates based on data taken at the given time, but analyzed using best available algorithms at publication time.  
\textbf{(a)}: Process infidelities of the three gates vs. wall time.
\textbf{(b)}: Diamond norm distance from estimated gates to targets, vs. wall time. Experiments from March 2015 surpass the best known diamond norm threshold of $6.7\times 10^{-4}$ with 95\% confidence, satisfying the threshold for fault-tolerance established in \cite{Aliferis:2009}.
\textbf{(c)}: Violation of Markovian model (in standard deviations $N_\sigma$) vs. wall time (see the section ``Quantifying non-Markovianity" for details).  Non-Markovian noise was progressively eliminated (e.g., by adding drift control and dynamical correction; see main text), guided by GST.}
\end{figure}
\begin{figure*}[tb]
\includegraphics[width= 1.0 \linewidth]{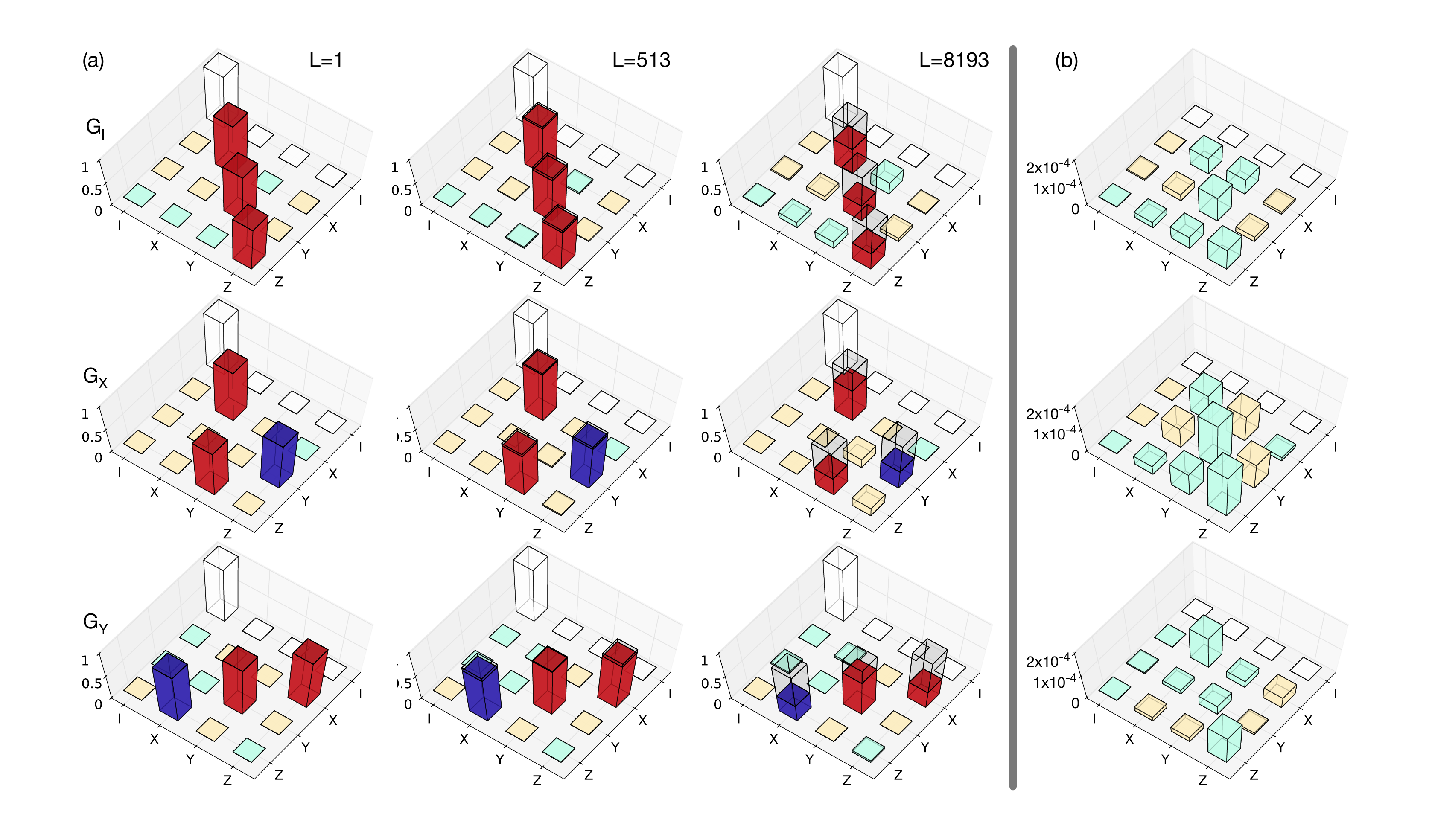}
\caption{\label{fig:GSTResults}
\textbf{Process matrices and error generators for the final gates implemented 30 March 2015.}
\textbf{(a)}:  GST estimates of the $G_I$, $G_x$ and $G_y$ gates, shown as superoperators in the basis of Pauli matrices, and based on data from gate sequences out to length 8192. For each estimate $\hat{G}_j$, we also show $\hat{G}_j^{513}$ and $\hat{G}_j^{8193}$ to emphasize errors. 
Bar height shows absolute value of matrix elements.  White bars are fixed by the TP (trace-preserving) constraint, red ones should (ideally) be $+1$, blue ones should be $-1$, and beige/teal ones should equal $0$ but are positive/negative, respectively.  Wireframes indicate the ideal (target) gates for comparison.
\textbf{(b)}: Error generators for each gate, using the same colors as (a). We define these as $ E = \ln \left(   G^{-1}_0  \hat { G}\right)$, where $\hat { G}$ is the estimate and $ G_0 $ is the target.
}
\end{figure*}

Using characterization procedures to debug and improve qubits has a long history.
A Ramsey fringe or Rabi oscillation experiment is a kind of limited tomography, combined with physical intuition, which is used to improve the quality of the quantum logic operations.
In typical tune-up procedures, different types of these experiments are iterated, until the qubit gates are deemed good enough to proceed.

Various improvements to this generic tuning-up scheme outlined above have been offered, including augmenting oscillation experiments to detect microwave pulse distortions \cite{Gustavsson:2013}, designing small sequences for error amplification \cite{Morton2005},
supplementing sequence experiments with RB to do detailed noise spectroscopy \cite{OMalley:2015}, or replacing them with iterative randomized benchmarking to guide the system toward higher RB fidelity operations \cite{Kelly:2014}.
All of these techniques, and other experiments combined with physical intuition, can be and have been used to produce qubits with very high-fidelity operations.  In comparison, GST has the distinct advantage in that it includes \noemph{all} experiments necessary for full and loophole-free qubit characterization.
It can be though of as systematic statistical inference on a provably sufficient set of Rabi/Ramsey experiments.

We used GST to analyze systematically and improve our trapped ion qubit operations over the course of five experimental runs from 17 April 2014 - 30 March 2015.  Experiments \#1-2 used the Sandia Thunderbird trap~\cite{stick_demonstration_2010}, and Experiments \#3-5 used Sandia's high-optical-access (HOA-2) trap.
Fig.~\ref{fig:resultsOverTime}a summarizes the gates' steady improvement over this period by tracking their process infidelities \cite{MagesanPRA11}, which corresponds to the RB error rate \cite{MagesanPRA2012}. 

Experiment \#1 detected severely non-Markovian behavior.  We sought to address this by stabilizing the microwave amplifier's temperature, and stabilizing microwave $\pi$-times using active feedback (drift control), as described in Methods.  Experiment \#2 showed improved fidelity in the $G_X$ and $G_Y$ gates, but no reduction in non-Markovianity.  We then moved our qubit to the HOA-2 trap, and improved trap stability.  In Experiment \#3, GST showed significant improvements in fidelity and Markovianity, and that $G_I$ remained worse than the other gates.  To improve it, we changed $G_I$ from ``do nothing for one clock cycle'' to the dynamical decoupling pulse sequence $X_\pi W_{1.25 \pi} (X)_{-\pi} W_{1.25 \pi}$, where $X_\pi$ and $Y_\pi$ denote $\pi$ rotations around $X$ and $Y$, respectively, and $W_{1.25 \pi}$ means ``wait for the duration of a $1.25\pi$-pulse''.  We also applied active drift control of the qubit frequency, and improved the calibration of the BB1 pulse sequences.  Experiment \#4 showed reduction of coherent errors in $G_X$ and $G_Y$, but persistent non-Markovian errors in $G_I$.  After we upgraded $G_I$ to the 2nd-order dynamical decoupling sequence $X_\pi Y_\pi X_\pi Y_\pi$~\cite{khodjasteh_dynamical_2009}, Experiment \#5 demonstrated uniformly excellent gates.  
Subsequent analysis indicates that the improved performance of $G_I$ stemmed largely from the constant duty cycle of the microwave system, rather than from the intrinsic properties of the decoupling sequence used.  This emphasizes that GST can identify specific errors, but not necessarily their cause.
The estimated process matrices for the gates are shown in Fig.~\ref{fig:GSTResults}a.  
Fig.~\ref{fig:GSTResults}b shows the error generators, defined as $  E = \ln \left(   G^{-1}_0  \hat { G}\right)$, where $\hat { G}$ is the estimate and $ G_0 $ is the target.
\subsection{Demonstrating suitability for fault tolerance}
Useful quantum computation is expected to require fault tolerant error correction.  The most important milestone for a quantum operation is, therefore, ``Is it suitable for use in FTQEC?'' Operations that induce too much error will cause FTQEC protocols to fail.  Demonstrating conclusively that gates are suitable for fault tolerance requires: (1) establishing a sufficient condition for the gates to not induce failure; and (2) showing that the gates satisfy that condition, with high confidence, by means of experimental data that are inconsistent with \noemph{all} gates that don't satisfy the condition.

Demonstrating suitability for fault tolerance using infidelity alone is hard.  Threshold theorems against general errors (arbitrary CP maps) are stated in terms of the diamond norm distance between the real and ideal gates \cite{aharonov:1998,Aharonov:2008},
\begin{equation}\label{eq:diamondNorm}
|| G - G_0 || _\diamond = \sup_{\rho} || (G \otimes \mathbbm{1}_d ) [\rho] - (G_0 \otimes \mathbbm{1}_d ) [\rho] ||_1,
\end{equation}
where $d$ is the system's Hilbert space dimension, $|| \cdot ||_1$ is the trace norm, and the supremum is over density matrices $\rho$ with dimension $d^2$ \cite{Benenti2010}.
Because the diamond norm error can be as large as $\sqrt{\epsilon_{\mathrm{RB}}}$ \cite{Kueng15,SandersNJP16}, even a spectacular RB result like $\epsilon_{\mathrm{RB}}=10^{-6}$ \cite{harty:2014} only establishes an upper bound of $10^{-3}$ on the diamond norm.  The best-known proof of fault tolerance against general noise \cite{Aliferis:2006} derived a threshold of $2.3\times 10^{-5}$ against stochastic noise, and generalized it to a diamond norm threshold of $\sim$$10^{-5}$ against general (coherent) noise. This was subsequently improved to $1.94\times10^{-4}$ \cite{aliferis:2007} and finally to $6.7\times10^{-4}$ \cite{Aliferis:2009}, the highest (currently) proven threshold against general noise.

Unlike RB, GST enables direct computation of the diamond norm between the estimated and target gates (we use a semidefinite program \cite{Watrous2012}). Fig.~\ref{fig:resultsOverTime}b shows the diamond norm error of our gates over time, culminating on 30 March 2015 in diamond norm error rates (with 95\% confidence intervals) of $(1.58 \pm 0.15)\times 10^{-4}$, $(1.39 \pm 0.22)\times 10^{-4}$, and $(1.62\pm 0.27)\times 10^{-4}$ for $G_I$, $G_X$, and $G_Y$ respectively.  All three gates surpass the threshold with 95\% confidence.  
(In point of fact, they surpass even the older $1.94\times10^{-4}$ threshold with 95\% confidence.)

We note that, although we only demonstrated Clifford operations, and non-Clifford operations are needed for universal control, FTQEC is possible only using Cliffords.
Furthermore, we can still extrapolate the performance of non-Clifford gates (e.g., a T gate) in our system.  
A pessimistic estimate of the error on an $X$ $\pi/4$ rotation, for example, would simply be the same as the error on the $X$ $\pi/2$ gate that we characterized.  
This is because implementing the $X$ $\pi/4$ gate in practice is equivalent to running the $X$ $\pi/2$ gate for a shorter duration.

\subsection{Quantifying non-Markovianity}
\label{sec:NonMarkov}
\begin{figure*}[tb]
\includegraphics[width= 1.0 \linewidth]{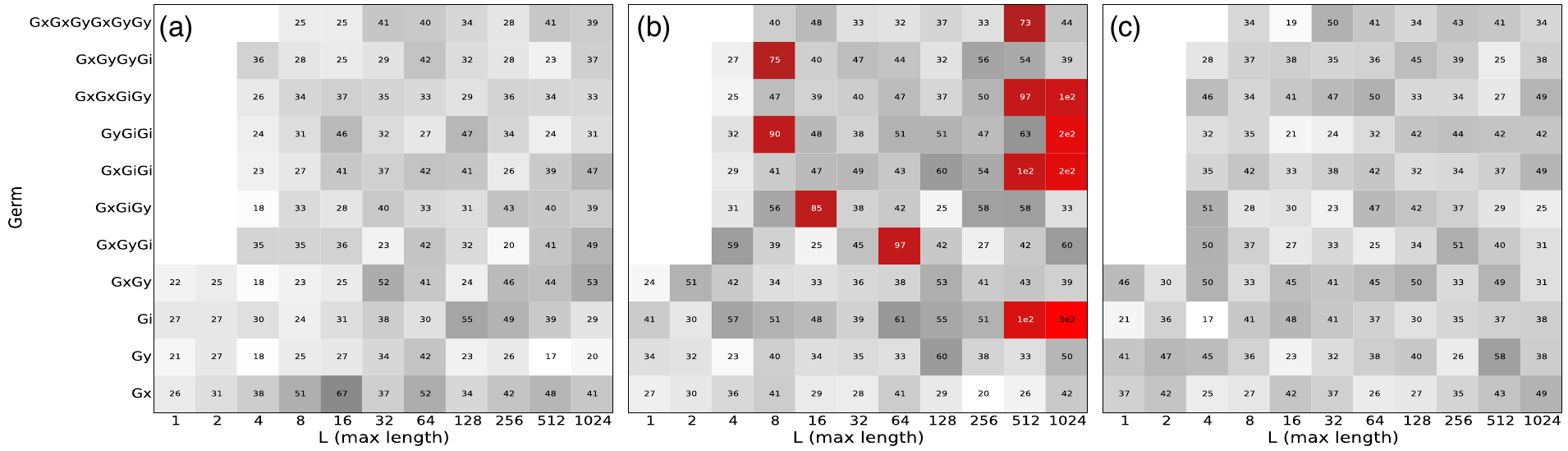}
\caption{\label{fig:LogLBoxes}
\textbf{$\logL$ box plots for gate set tomography fits for three datasets.}
In each individual box, $2\Delta\logL$ is summed over a set of 36 different gate strings.  If the underlying system is Markovian, $2\Delta\logL$ is (approximately) a $\chi^2_{36}$ random variable.  
The box color indicates the $2\Delta\logL$ score for that collection of sequences.  
Grey indicates score values that are expected due to statistical fluctuations, while red indicates significant model violation at 95\% confidence.
That is,  if the gates are Markovian, the probability of observing even one red square is at most 5\%.
Note that certain germs are too long to appear at $L=1,2,4$.
\textbf{(a)}:  $\logL$ box plot for simulated Markovian data.  
\textbf{(b)}:  $\logL$ box plot for experimental data from  2 March, 2015, with indications of strong non-Markovianity.
\textbf{(c)}:  $\logL$ box plot for experimental data from 30 March, 2015, with vastly decreased non-Markovianity.}
\end{figure*}

In real experimental systems, repeated quantum operations are never actually identical.  For example, experimental imperfections in the stability of the system may cause quantum operations to drift over time.  Collectively, we refer to all such non-repeatability as \noemph{non-Markovianity}.  
It represents a significant potential problem for fault tolerance, as proofs of fault tolerance thresholds are typically carried out using Markovian error models.  So, to be confident that a gate set is suitable for FTQEC (i.e., achieves a fault tolerance threshold), we would like to demonstrate that non-Markovian behavior is absent.  This is not feasible, for two reasons.  First, all physical systems (including qubits) are at least a tiny bit non-Markovian.  Second, ``non-Markovian noise'' is so general that there is always some conceivable mechanism that would elude detection by any protocol (not just GST).  Our goal is to reduce detectable non-Markovian behavior to the point where its visible effects are consistent with the FT threshold.

We use GST results to debug non-Markovian effects and achieve this goal, as illustrated in Fig.~\ref{fig:resultsOverTime}c.  Doing this is nontrivial, because neither GST nor process tomography is actually designed to characterize non-Markovianity.  In GST's underlying model, the qubit is Markovian:  its state at time $t+1$ is determined \noemph{completely} by (1) its state at time $t$ and (2) the operation applied at time $t$.  This assumption is far reaching.  It implies that noise in the logic gates is stationary, uncorrelated in time, memoryless, and independent of context (e.g., what gates were recently applied).  It implies that the gate operations can (for a single qubit) be represented as static $4\times 4$ superoperators, and that state preparation and measurements may each be represented as static four-dimensional vectors and dual vectors (respectively), in Hilbert-Schmidt space.

As indicated above, the Markovian assumption is not strictly true for any experimental system.  In addition to slow drift, there may be correlations between errors in consecutive gates, and the ``qubit'' may not even be a two level system (e.g., due to leakage levels).  These are all examples of non-Markovianity, and lie outside the GST model.

In principle, all guarantees about GST are void in the presence of non-Markovian noise, as there are no process matrices to measure or report.  However, for many typical non-Markovian behaviors, GST degrades in a quantifiable way.  These kinds of non-Markovian noise cause data that are consistent with \noemph{no} Markovian gate set, and this failure to fit the data can be quantified.  Since data generated by any Markovian model could be fit with predictable accuracy, significant badness-of-fit can be interpreted as violation of the model and therefore as non-Markovianity, though the particular type cannot be easily identified.  As long as the data appear \noemph{sufficiently} Markovian, the GST estimate will be fairly reliable and have significant predictive power.  

To quantify non-Markovianity, we consider the log-likelihood (Eq. \ref{eqGateStringLogL})
\begin{equation}
\logL = N\sum_s\left({f_s \log(p_s) + (1-f_s) \log(1-p_s)}\right),
\end{equation}
where $f_s = n_s/N.$ The best \noemph{conceivable} fit to a dataset would be one where $p_s = f_s$ for every sequence $s$.  Thus, the \noemph{entropy} of a dataset is an upper bound on $\logL$,
\begin{multline}
\logL \leq NH(\{f_s\}) \equiv \\N\sum_s\left({f_s \log(f_s) + (1-f_s) \log(1-f_s)}\right).
\end{multline}
We define the quantity $\Delta\logL=NH(\{f_s\})-\log \mathcal L_\text{max}$.

Standard properties of maximum likelihood estimation theory \cite{WassermanBook13} imply that if
\begin{enumerate}
\item the data were in fact generated by some gate set,
\item there are $N_p$ free parameters in the gate set, and
\item the dataset contains $N_s > N_p$ distinct gate sequences
\end{enumerate}
then as $N\to\infty$, $2\Delta\logL$ is a random variable with a $\chi^2_{k}$ distribution, where $k = N_s - N_p.$
This means that its expected value is $\expec{\chi^2_k}=k$, and its RMS variance is $\pm\sqrt{2k}$.  
Thus, if the fit is ``good'', then $2 \Delta\logL$ should lie roughly within the interval $[k-\sqrt{2k},k+\sqrt{2k}]$. Hence, by comparing the difference $2\Delta\logL - k$ to $\sqrt{2k}$ we can determine how well the Markovian model was able to fit the data.

We quantify goodness-of-fit by $N_\sigma$, the number of standard deviations from the expected mean the expected mean the log-likelihood score is:

\begin{equation}
N_\sigma = \frac{2\Delta\logL-k}{\sqrt{2k}}.
\end{equation}

We can also calculate $2\Delta\logL$ for individual experiments or subsets of gate sequences.  Figure~\ref{fig:LogLBoxes} illustrates this, where $2\Delta\logL$ is shown for every individual experiment associated with each power of each germ (for a total 36 experiments per collection, due to six preparation fiducials and six measurement fiducials).  This analysis makes it possible to see whether non-Markovianity increases with sequence length (it usually does, because longer sequences amplify slowly varying noise), and which sequences are particularly inconsistent with the best Markovian fit.

Figure~\ref{fig:LogLBoxes} compares $2\Delta\logL$ scores for a simulated (perfectly Markovian) dataset to those for two experimental data sets, one from 02 March 2015, and the other from 30 March 2015.  The March 2 experimental dataset is highly non-Markovian, while the March 30 dataset looks very similar to the simulated dataset.  These data demonstrate the degree to which we are able to stabilize our qubit and reduce non-Markovian effects.  In the final run on March 30, Markovianity is violated only at the $4\sigma$ level.  While this is statistically significant -- it implies with high confidence that the gates are not perfectly Markovian -- it is not practically significant.  To see this, recall that this is an extraordinary sensitive experiment, as witnessed by the fact that the error bars on the diamond norm are $\pm2\times 10^{-5}$.  This sensitivity extends to non-Markovian behavior as well.  Reducing the sensitivity of the experiment by a factor of 4 (either by reducing maximum $L$ by a factor of 4, or by reducing $N$ by a factor of 16) would render the non-Markovianity undetectable, at the cost of increasing the error bars by a factor of 4 to $\pm8\times 10^{-5}$.  This implies that the observed non-Markovianity is effectively equivalent to less than $10^{-4}$ additional diamond norm error, which is comfortably below the threshold.

\subsection{Comparison to randomized benchmarking}
\label{subsec:RB}
As of this writing, randomized benchmarking (RB) is the de facto standard in qubit characterization.  As a consistency check, we perform RB simultaneously with the final GST experiment (by interleaving the GST and RB sequences over the entire period of experimentation), to see whether GST correctly predicted the results of RB.  We follow the experimental and analysis procedure of Ref.~\cite{wallman_randomized_2014}, and use RB sequences ranging in length from 2 gates to 1970 gates.

Strictly speaking, RB measures the error rate per Clifford operation.  Our Clifford operations are, as is usual, compiled into elementary $\{G_i,G_x,G_y\}$ gates, with an average of $3.125$ elementary gates per Clifford.  Analysis of the data in strict accordance with the literature (i.e., plotting survival probability versus \# of Cliffords in the sequence) yielded an experimental error rate of $(1.65\pm0.03)\cdot10^{-4}$ per Clifford operation.  Dividing this by 3.125 (a questionable but common practice) suggests a per-gate error rate of about $\epsilon_{\mathrm{RB}} = (5.28\pm0.10)\cdot10^{-5}$.

\begin{figure}[tbh!]
\includegraphics[width= 1.0 \linewidth]{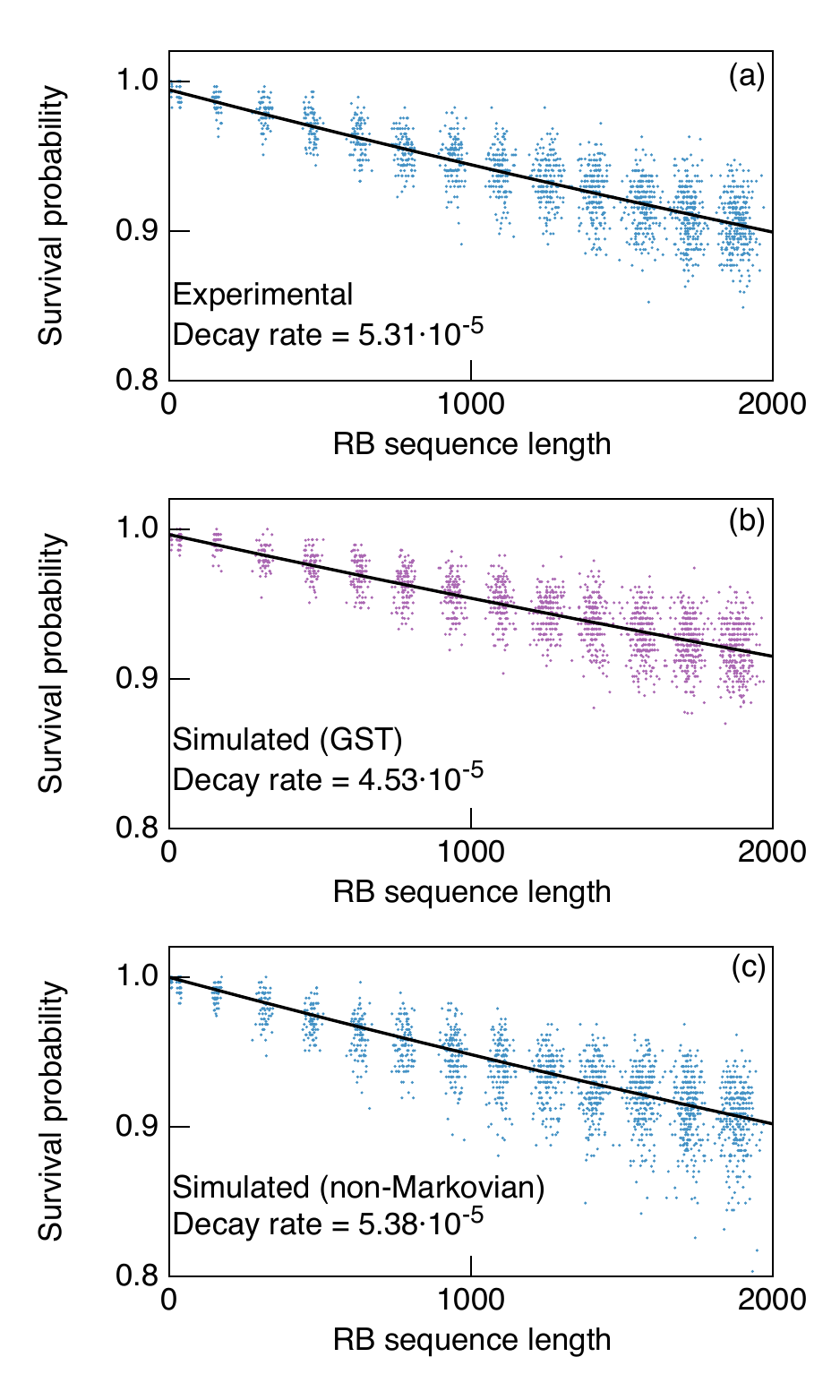}
\caption{\label{fig:SuppMat-RB}  \textbf{Randomized benchmarking results.}
Colored dots are experimental or simulated data points; lines are exponential decay fits to the data.  
\textbf{(a)}: Experimental RB data. 
\textbf{(b)}: RB data simulated using the gate set $\mathcal{G}_{0}$ derived from experimental GST results. 
\textbf{(c)}: RB data simulated using the non-Markovian gate set $\mathcal{G}_{comp}$.  
Here, $\mathcal{G}_{comp}$ is generated using the composite Nyquist-limited noisy gate set model proposed in the section ``Comparison to Randomized Benchmarking.''
This model toggles between slightly over- and under-rotated gates with every gate application,
which is a reasonable scenario for our qubit due to our use of dynamically corrected gates.
The experimental RB decay rate is $(5.31\pm0.16)\cdot10^{-5}$, which is indistinguishable from $\mathcal{G}_{comp}$'s RB decay rate of $(5.38\pm0.17)\cdot10^{-5}$, but distinct from $\mathcal{G}_0$'s RB decay rate of $(4.53\pm0.25)\cdot10^{-5}$,
demonstrating the plausibility of our non-Markovian model in explaining the apparent discrepancy between experimental RB and GST.}
\end{figure}

However, our main goal is to compare the RB data with GST's predictions for it.  For this purpose, we find it more informative to fit (and plot; see Fig.~\ref{fig:SuppMat-RB}) the observed probabilities versus the number of elementary gates in the sequence.  All the rest of the analysis in this section is based on this analysis method, which yields a per-elementary gate RB error rate of $\epsilon_{\mathrm{RB}} = (5.31\pm0.16)\cdot10^{-5}$.  Error bars are 95\% confidence intervals.  The experimental error bars are calculated via non-parametric bootstrap (by resampling the experimental data with replacement).  We then simulate those RB experiments using the GST estimates.  The GST results predict an RB error rate of $\epsilon_{\mathrm{RB}} = (4.53\pm0.25)\cdot10^{-5}$ (see Figs.~\ref{fig:SuppMat-RB}a and~\ref{fig:SuppMat-RB}b).  The simulated error bars are calculated via parametric bootstrap.  (The GST estimate is used to generate many sets of simulated GST experiments, each of which in turn yields a new GST gate set estimate.  This ensemble of estimates then generates an ensemble of simulated RB decay rates, from which the simulated error bars are derived.)

While these decay rates are nearly identical, there is a statistically significant discrepancy.  The most obvious explanation is a flaw in the GST analysis, but we find that extensive simulations with known Markovian gates rule this possibility out.  We believe that the discrepancy stems from physical causes -- i.e., from non-Markovian noise.  The most common form of non-Markovian errors is low-frequency drift, which manifests in both RB and GST as coherent errors that remain nearly fixed over the course of any one sequence, but change from sequence to sequence (and between repetitions of a single sequence).  In the presence of such effects, GST typically \noemph{overestimates} the RB decay rate, because GST amplifies coherent errors to which RB sequences are relatively insensitive (Markovian or not).  Thus, GST typically reports a higher rate of Markovian noise in a quixotic attempt to fit its data, while RB simply doesn't see the noise.  

But in this experiment, we observe the opposite effect.  Instead of \noemph{over}estimating the RB error rate, GST \noemph{under}estimates it.  While the exact cause remains uncertain, we observe that this behavior is completely consistent with anti-correlated noise (each gate flips between under- and over-rotation at each application) induced by dynamically corrected gates \cite{khodjasteh_dynamical_2009}.

Here is a concrete model that reproduces this behavior:  Consider a unitary error that varies in time -- but instead of varying slowly, it oscillates at the system's Nyquist frequency (i.e., flips sign every clock cycle).  For simple gates implemented with a single pulse, this would be highly implausible.  In this experiment, however, we implement dynamically corrected gates (DCG).  The simplest DCG is a dynamically corrected idle gate (our $G_I$).  This is nothing but dynamical decoupling -- periodic $X_\pi$ pulses that echo away small $Z$ rotations.  Such sequences create a ``toggling frame'' for the qubit that flips sign twice per clock cycle.  Any timing or amplitude errors in the pulses can leave a residual error that flips sign every clock cycle, making a plausible noise model for a DCG.

We model this effect by augmenting the qubit state space with a classical binary variable $q\in\{-1,+1\}$.  We define a composite gate set $\mathcal{G}_{comp}$, based on a standard gate set $\mathcal{G}_0$, which consists of two single-qubit gate sets $\mathcal{G}_+$ and $\mathcal{G}_-$ that act \noemph{conditionally} on the value of the classical bit $q$, which flips every time a gate is applied.  These gate sets are identical to $\mathcal{G}_0$, except that the $G_X$ and $G_Y$ elements of $\mathcal{G}_+$ have a fixed, slight \noemph{over}-rotation by an angle $\theta$, while the $G_X$ and $G_Y$ elements of $\mathcal{G}_-$ have a fixed, slight \noemph{under}-rotation by $\theta$.  At the beginning of each simulated experiment, $q$ is chosen randomly.

$\mathcal{G}_{comp}$ acts on an $8$-dimensional state space, and data generated from it is not fully consistent with any Markovian single-qubit gate set.  But GST can be applied to that data, and will find the Markovian single-qubit gate set that fits it best.  (Indeed, as no experimental system is perfectly Markovian, this is in essence what GST always does.)

We generate simulated data, with finite sample error, for all the GST and RB experiments performed on 30 March, 2015.  For this simulation, $\mathcal{G}_{comp}$ is defined by setting $\mathcal{G}_0$ equal to the GST estimate from 30 March, 2015, and setting $\theta=1.25\cdot10^{-2}$.

Analyzing the GST data generated by $\mathcal{G}_{comp}$ yields an estimated gate set almost identical to that obtained from experimental data ($\mathcal{G}_0$). All but two of the 36 free gate matrix elements are within the 95\% confidence intervals assigned to $\mathcal{G}_0$, and the three remaining elements are at most $0.05\sigma$ outside them.  Every gate is within $4.4\cdot10^{-5}$ (in diamond norm) from the corresponding gate in $\mathcal{G}_0$.  We conclude that GST, as performed, cannot distinguish the composite model ($\mathcal{G}_{comp}$) from $\mathcal{G}_0$.

The RB data simulated with $\mathcal{G}_{comp}$ also matches the experimental RB data almost perfectly, yielding an RB error rate of $(5.38\pm0.17)\cdot10^{-5}$ that is statistically indistinguishable from the experimentally observed RB decay rate of $(5.31\pm0.16)\cdot10^{-5}$.  Both datasets (experimental and simulated-by-$\mathcal{G}_{comp}$) are shown in Figure~\ref{fig:SuppMat-RB}, along with RB data simulated from $\mathcal{G}_0$.  We conclude that RB observes significantly different error rates for $\mathcal{G}_0$ and $\mathcal{G}_{comp}$.

This does not imply that our qubit really is described by $\mathcal{G}_{comp}$, but it demonstrate a plausible non-Markovian model that is fully consistent with our data.  There might be many other (different) non-Markovian models equally consistent with it.  And while certain kinds of non-Markovian noise can be detected by RB \cite{Ball2016} and GST, neither GST nor RB are designed to function reliably in the presence of \noemph{any} non-Markovian noise, so neither of them is explicitly ``right'' or ``wrong'' for this case.

\subsection{The relative power of RB and GST}

RB and GST share the common framework of data from gate sequences (circuits) that are (1) diverse, (2) repeated, and (3) long.  But they are distinguished by the \noemph{kind} of sequences performed.  RB sequences are random, for the specific purpose of ``twirling'' the noise.  GST sequences are structured and periodic, for the specific purpose of amplifying errors.

This difference is fundamental.  It makes RB intrinsically insensitive to coherent errors, which dominate the diamond norm error metric \cite{SandersNJP16,Kueng15}.  For example, suppose that one logic gate over-rotates by a small angle $\theta$, while the others are perfect.  In random sequences containing $L$ applications of this gate, it will (by construction) be interleaved with other gates chosen randomly.  The rotations by $\theta$ will add up incoherently, producing (on average) a total rotation of $\theta\sqrt{L}$, and therefore an error probability of $L\theta^2$.  Thus, a coherent error by $\theta$ appears (in RB) as an incoherent error of probability $\theta^2$.

But circuits of practical interest are not random.  Since not all ``useful'' circuits are known at this time, it is wise to consider how errors affect arbitrary circuits in the \noemph{worst} (most fragile) case.  For the example given above, the worst case is a periodic sequence in which the imperfect gate appears $L$ consecutive times.  Rotations add up coherently, the final angle is $L\theta$, and the final error probability is $L^2\theta^2$.  So, for example, a $10^{-3}$ rotation can cause a $1\%$ failure rate after just $L=100$ gates.  In randomized circuits, the same failure rate would require $L=10^4$ gates.

The diamond norm metric is a strict upper bound on the rate at which failure probabilities can grow, and so it takes account (by construction) of the worst-case behavior given above.  The diamond norm error for a small coherent error by angle $\theta$ is $O(\theta)$.  Process infidelity (closely related to the RB error rate) does not account for worst-case behavior, and the process infidelity for a small coherent error by angle $\theta$ is $O(\theta^2)$.

GST intentionally implements a wide variety of periodic sequences, to ensure that at least one of them is approximately ``worst case'' for every possible coherent error.  This allows GST to detect coherent errors of size $\theta$ using sequences of length $L = O(1/\theta)$, repeated $O(1)$ times.  Detecting the same error with randomized sequences requires much long sequences of length $L = O(1/\theta^2)$, or else a much higher number of repetitions (both of which correspond to orders of magnitude more time and effort).

Periodic sequences might be incorporated into RB, to make it more sensitive.  Doing so, however, would eliminate its characteristic feature.  Such a protocol would no longer be \noemph{randomized} benchmarking.  On the other hand, there are several interesting variations of RB that retain its randomized nature, most notably interleaved benchmarking \cite{MagesanPRL12}, RB tomography \cite{kimmel13}, and unitarity benchmarking (URB) \cite{WallmanNJP15}.  While interesting in their own right, they are all subject to the same trouble:  random gate sequences are much less sensitive to coherent errors than periodic ones, and therefore every form of RB is necessarily inefficient at detecting coherent errors.

Unitarity benchmarking is particularly interesting, since (unlike other forms of RB) it can separate coherent and incoherent errors, and therefore provide good information about diamond norm error rates.  Unfortunately, it is (compared with GST) extremely inefficient at doing so.

Wallman et al. \cite{WallmanNJP15} defined a quantity $u$ (unitarity), which measures the rate of purity decay.  They gave an RB-like protocol for measuring it, and pointed out that $u$ and $r$ together could be used to bound the diamond norm error.  If
\begin{equation}
u=u_{\mathrm{min}} \equiv \left(1-\frac{d}{d-1}r\right)^2,
\end{equation}
then the noise is purely incoherent, and the diamond norm error is $O(r)$.  If $u-u_{\mathrm{min}}$ is sufficiently small, then the errors are primarily incoherent, and the diamond norm error remains $O(r)$.

However, the actual bounds (see \cite{Wallman15FT}) are of the form
\begin{equation}
\vert\vert\cdot \vert\vert_{\diamond} = O(\sqrt{u-u_{\mathrm{min}}}).
\end{equation}
We have demonstrated that our gates' diamond norm error is $O(r)$ using GST.  Doing the same thing using unitarity requires showing that $u-u_{\mathrm{min}} = O(r^2)$.  But $1-u$ is itself an RB-type quantity, meaning that it appears as an error rate (in experiments that measure purity), and is measured using RB.  As a result, showing that $u-u_{\mathrm{min}} = O(r^2)$ is equivalent to:
\begin{enumerate}
\item Performing standard RB to measure $r$, the decay rate of sequence fidelity.
\item Performing a different RB-like experiment to measure $r' = 1-u$.
\item Demonstrating (based on those experiments) that $r-r'= O(r^2)$.
\end{enumerate}
For $r=10^{-4}$ (the regime we access experimentally), this requires measuring both $r$ and $r'$ to $10^{-8}$ precision.  This is extraordinarily hard.  The most efficient way to do it is using sequences of length $L \approx 10^4$.  These would yield survival probabilities around $p_L \approx 1/e$.  Achieving the necessary precision would require estimating $p_L$ to $\pm 10^{-4}$, which would require approximately $N=10^{8}$ repetitions (because the uncertainty is $O(1/\sqrt{N})$).  This is at least $10^6$ times more repetitions than would be required for standard RB, or for GST, and is completely impractical.

\subsection{Validating $10^{-5}$ accuracy with simulations}
\begin{figure}[tbh!]
\includegraphics[width= 1.0 \linewidth]{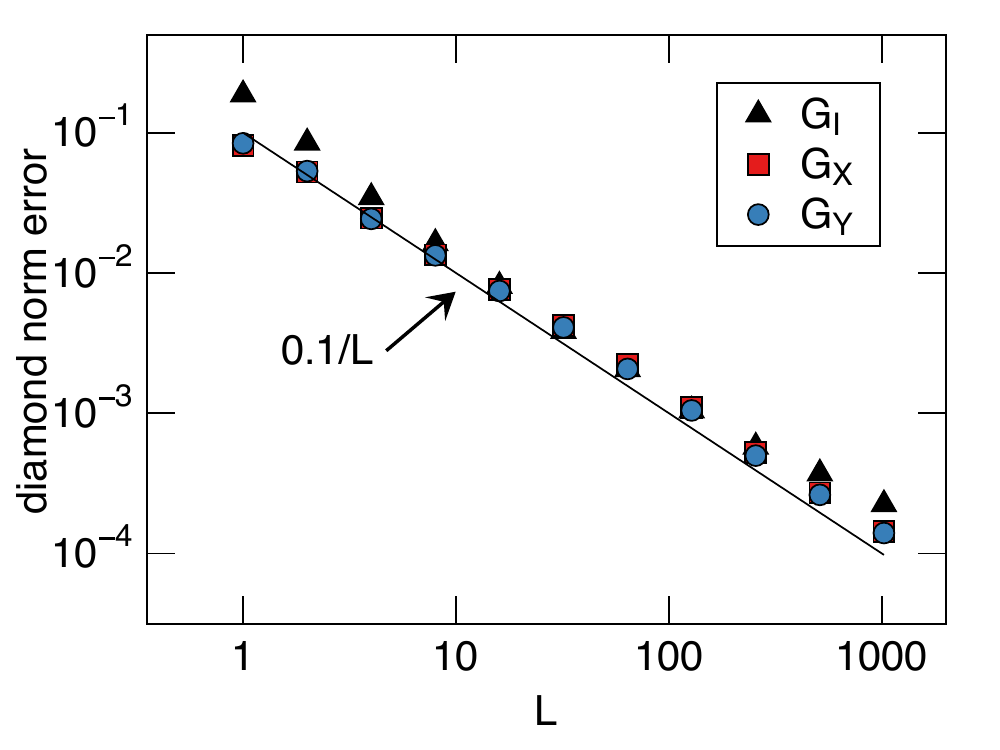}
\caption{\label{fig:Heisenberg}
\textbf{Confirmation of error scaling in gate set tomography.}
Here, we show diamond norm difference between true and estimated gates in \noemph{simulated} GST with small unitary errors.
Mean diamond norms are shown, averaged over 100 trials.  Estimation error scales as $1/L$, where $L$ is the maximum sequence length in the data.  
Each trial uses $N=50$ samples per experiment.}
\end{figure}

We have claimed uncertainties (error bars / confidence regions --- see Methods) of about $10^{-5}$ for diamond norms and process matrix elements.  This is remarkable, and demands supporting evidence.
To confirm this behavior we simulate GST experiments using (known) gate sets with unitary errors.  The results (Fig.~\ref{fig:Heisenberg}) confirm Heisenberg scaling: diamond norm distance between estimated and true gates decreases with the maximum sequence length ($L$) as $1/L$.  This scaling holds up to $L\approx 1/\epsilon$, where $\epsilon$ is the stochastic error rate.  This is consistent with the $\pm\sim10^{-5}$ observed error bars on diamond norm errors in our final experiment, for which $L=8192$.

\section{Discussion}

Gate set tomography allows us to achieve high-quality gates in a trapped-Yb$^+$-ion qubit, and to characterize it to unprecedented precision.
Although lower RB error rates have been reported in trapped-ion qubits \cite{BrownPRA11,harty:2014}, our gates are the first to demonstrably surpass a rigorous fault tolerance threshold against general noise.  GST is the first protocol that can efficiently demonstrate this important milestone \noemph{and} provide reliable feedback to debug and improve those gates.

Low-error single-qubit gates are just one of several critical achievements required to enable fault tolerant quantum computing.  Thus, this is only a first step.  But GST -- which can be generalized to 2-qubit gates and measurements -- does answer one key and pressing question:  ``Once suitable operations have been achieved, how can their performance be verified for a critical, objective observer?''  Randomized benchmarking can provide reliable information about process fidelity (which unambiguously captures stochastic or incoherent errors), but as of this writing, process fidelity is not known to be the relevant metric for fault tolerance.  An exciting recent development in this area is the introduction of randomized compiling \cite{Wallman15n}, which has the potential to provably reduce the importance of coherent errors.  But until and unless such techniques lead to a fault tolerance proof that is insensitive to them, and are confirmed to be practical in the context of FTQEC, coherent errors remain a point of concern.  GST provides an efficient way to diagnose and bound \noemph{all} Markovian errors in gates.

\section{Methods}

\subsection{Experimental details}

In Sandia's Thunderbird trap, ions were trapped $80\operatorname{\mu m}$ above the trap surface. Typical trap frequencies were $0.5\operatorname{MHz}$, $1.8\operatorname{MHz}$, and $2.3\operatorname{MHz}$, for the axial and two radial modes, respectively. In the HOA-2 trap, ions were trapped $68\operatorname{\mu m}$ above the trap surface and trap frequencies of $0.5\operatorname{MHz}$, $2.2\operatorname{MHz}$, and $2.8\operatorname{MHz}$ were achieved. Typical trapping times were several hours for the Thunderbird trap and up to $100\operatorname{h}$ for the HOA-2 trap. Coherence times were measured to be $\sim$$1\operatorname{s}$ in both traps, and were most likely clock-limited.

The microwave radiation used for qubit manipulation was generated by single-sideband (SSB) modulating the output of a $12.600\operatorname{GHz}$ dielectric resonator oscillator (DRO) with the output of a direct digital synthesizer (DDS) near $42.812\operatorname{MHz}$. The master clock for the DRO and DDS was generated by a rubidium frequency standard. The output of the SSB modulator was amplified and directed parallel to the trap surface using a microwave horn. The microwave frequency and phase was controlled via the DDS and approximately square pulses were generated by switching the output of the DDS using a high-isolation rf switch. An offset was added to the constituent pulses of the BB1 pulse sequence to compensate for switching imperfections.

Drift control of the microwave $\pi$-time was realized by interleaving experiments in which the ion was initialized, exposed to a $10.5 \pi$ microwave pulse, and measured.
The $\pi$-time was adjusted after state detection; upon seeing $\ket{1}$, $\pi$-time was decreased by $0.625\operatorname{ns}$, while upon seeing $\ket{0}$, $\pi$-time was increased by $0.625\operatorname{ns}$. 
For the next experiment, the $\pi$-time was then truncated to the time resolution of the experimental control ($5\operatorname{ns})$. Similarly, drift control of the qubit frequency was implemented by interleaving a Ramsey experiment in which the ion, after state initialization, is subject to: (1) a $G_X$ gate, (2) a $25\operatorname{ms}$ wait time, and (3) a $G_Y$ gate. 
Upon state detection, the qubit frequency was adjusted by $+8\operatorname{mHz}$ for a $\ket{1}$ result, and by $-8\operatorname{mHz}$ for a $\ket{0}$ result.

\subsection{Linear GST}
Linear-inversion GST (LGST) is a highly reliable but low-accuracy way to obtain an initial estimate of the gate set that serves as a seed for further refinement by long-sequence GST (see next subsection).  LGST is essentially simultaneous ``uncalibrated'' process and state tomography.  By performing process tomography-like experiments on a set of gates, as well as the null operation (i.e., the ``do nothing for no time'' operation), LGST can provide rough estimates of all the gates involved, as well as the state preparation and measurement operations.  LGST requires minimal assumptions about the various operations (unlike standard tomography), and computes its estimates using only basic linear algebra (the most complicated step is matrix inversion).  A detailed explanation of the LGST procedure is provided in \cite{GST2013}; LGST is also described in \cite{Greenbaum15}.

\subsection{Analyzing long sequences in GST}
\label{sec:LSGST}
GST incorporates data from long sequences in two stages.  The first stage consists of several iterations, each of which performs a minimum-$\chi^2$ estimation.   Each iteration takes the result of the previous iteration as a seed, and includes successively more of the long-sequence data.  The second stage is a maximum-likelihood estimation which is seeded from the first stage and uses all of the data. This procedure consistently avoids local minima in the objective function.  In this section, we give the details of this algorithm (outlined in Fig.~1).

The iterative fitting procedure starts by fitting only data from the shortest gate sequences (which are easy to fit \noemph{and} insensitive to most non-Markovian noise), then successively adds longer and longer sequences (with base sequence length $L\leq 1,2,4,8,\ldots$).
Since we get an estimate at each intermediate $L$, it is possible to quantify not just the goodness of the \noemph{best} fit, but how the goodness-of-fit behaves as longer and longer sequences are added in, which is useful for debugging.

At each step in the iterative process, we vary the gate set to minimize Pearson's $\chi^2$ test statistic, which measures the discrepancy between a predicted 
probability $(p)$ and an observed frequency $(f)$.  It is defined as
\begin{equation}
\chi^2 = N\frac{(p-f)^2}{p},
\end{equation}
where $N$ is the number of samples taken.  In this analysis, $\chi^2$ is used to compare the set of probabilities predicted by a gate set ($p_s$) and the frequencies obtained from a dataset ($f_s$).  Each experiment (i.e.~gate sequence) $s$ is associated to two probabilities:  ``plus'' has probability $p_s$ and ``minus'' has probability $1-p_s$.  The $\chi^2$ of a single gate string $s$ is
\begin{equation}
\chi^2_s = N\frac{(p_s-f_s)^2}{p_s} + N\frac{(p_s-f_s)^2}{1-p_s} = \frac{N(p_s-f_s)^2}{p_s(1-p_s)},\label{eqGateStringChi2}
\end{equation}
where $N$ is the number of times the experiment $s$ was performed, $p_s$ is the probability of a ``plus'' outcome as predicted by the gate set, and $f_s$ is the observed frequency of ``plus''.  The total $\chi^2$ for a dataset $\mathcal{S}$ is just the sum 
\begin{equation}
\chi^2_\mathcal{S} = \sum_{s\in\mathcal{S}}{ \chi^2_s}
\end{equation}
To estimate our gate set parameters, we minimize $\chi^2_\mathcal{S} $ at each iteration using the Levenberg-Marquardt algorithim implemented in \emph{SciPy} \cite{scipy}.

The final stage in long-sequence GST analysis is a maximum-likelihood estimation (MLE), based on numerical optimization of the log-likelihood function  $\logL$.
The log-likelihood for an $n$-outcome system with predicted probabilities $p_i$ and observed frequencies $f_i$ ($i=1\ldots n$) is given by:
\begin{equation}
\logL = \sum_i N f_i \log(p_i).
\end{equation}
where $N$ is the total number of counts. 
Like the $\chi^2$ statistic, $\logL$ is used to 
compare the set of probabilities predicted by a gate set ($p_s$) to the frequencies obtained from a dataset ($f_s$).  
Each experiment (i.e.~gate sequence) $s$ is associated to two probabilities:  ``plus'' has probability $p_s$ and ``minus'' has probability $1-p_s$.  The $\logL$ contribution of a single gate string $s$ is
\begin{equation}
\logL_s = N f_s \log(p_s) + N (1-f_s) \log(1-p_s),\label{eqGateStringLogL}
\end{equation}
where $N$ is the number of times the experiment $s$ was performed, $p_s$ is the probability of a ``plus'' outcome as predicted by the gate set, and $f_s$ is the observed frequency of ``plus''.  The total log-likelihood for an entire dataset is just the sum 
\begin{equation}
\logL_\mathcal{S} = \sum_{s\in\mathcal{S}}{ \logL_s}.\label{eqDatasetLogL}
\end{equation}

We find the maximum of this quantity using the same Levenberg-Marquardt algorithm as above, in order to compute the final (modulo gauge optimization) estimate of the gates.

LGST would be a perfect estimator in the absence of finite-sample error.  However, it is inefficient with respect to accuracy.  Like process tomography, its inaccuracy scales as $O(1/\sqrt{N})$, which means that achieving $10^{-5}$ error bars on all parameters would require around $N=10^{10}$ repetitions of each experiment.  Long sequences amplify errors proportional to $L$, enabling inaccuracy of $O(1/L\sqrt{N})$ for all parameters.  (This scaling breaks down for $L\geq 1/\epsilon$, where $\epsilon$ is the rate of stochastic decoherence.  In our experiments, $\epsilon \leq 10^{-4}$, and we perform experiments as long as $L=8192 \approx 10^4$.)

We use a hybrid algorithm (involving both min-$\chi^2$ and MLE) because each of its components have certain weaknesses.  Empirically, we find that MLE is statistically well-motivated and avoids any bias, whereas $\chi^2$ optimization is numerically more stable and faster computationally but yields biased estimators, especially for the SPAM parameters.  
Our hybrid method combines both virtues, by using the more efficient and reliable min-$\chi^2$ algorithm to get a very good seed for the final (unbiased) MLE.

On a modern laptop, single-qubit GST with maximum $L=1024$ can run in under 1 minute; the analysis for maximum $L=8192$ takes about 40 minutes. 

\subsection{Selecting gate sequences for GST}

The data that GST use to reconstruct a gate set come from performing \noemph{gate sequences} (\noemph{i.e.} quantum circuits).  Every gate sequence necessarily comprises (i) initialization, (ii) some gates, and (iii) measurement (which yields a \noemph{count} that is recorded in the dataset).  The sequences used for GST have an additional structure:
\begin{enumerate}
\item Each GST sequence begins with a \noemph{preparation fiducial sequence}, and ends with a \noemph{measurement fiducial sequence}, with an ``operation of interest'' sandwiched in the middle.
\item The ``operation of interest'', which could in principle be any gate sequence, is chosen to be a \noemph{germ power sequence} -- i.e., a short ``germ'' sequence, repeated an integer number of times.
\end{enumerate}
Thus, every GST sequence is of the form $F_j g_k^L F_i$, where $F_i$ and $F_j$ are preparation and measurement fiducials (respectively), $g_k$ is a germ, and $L$ is an integer.  $F_i$ and $F_j$ range exhaustively over a set of 6 fiducial sequences, while $g_k$ ranges exhaustively over a set of 11 germs.  In this section, we explain how the fiducials and germs are chosen.

The purpose of the fiducials is to prepare a sufficiently diverse set of input states and measurements to completely probe the operation of interest.  This is achieved if (and only if) the input states $\{\rho_i\}\equiv  \{F_i\sket{\rho}\}$ and the measurement effects $\{E_j\} \equiv \{\sbra{E}F_j\}$ are both \noemph{informationally complete} (IC).  A set of matrices is IC if and only if it spans the vector space $\mathcal{B}(H)$ of matrices.  This requires at least $d^2$ linearly independent elements.  

In general, any randomly chosen set of $d^2$ states or effects will be IC.  So, for single-qubit GST, we could choose $d^2=4$ random fiducial sequences.  However, while the resulting $\{\rho_i\}$ and $\{E_j\}$ will almost certainly be linearly independent, they may be \noemph{close} to linearly dependent.  This property is quantified by the spectrum of the Gram matrix $\tilde{\Id}$, defined by
\begin{equation}
\tilde{\Id}_{j,i} = \sbraket{E_j}{\rho_i}.
\end{equation}
If either set fails to be IC, the Gram matrix will fail to have $d^2$ non-zero (to machine precision) singular values.  As any one of the $d^2$ largest singular values becomes small, inverting the Gram matrix on its support (as is required for LGST) becomes ill-conditioned, and finite-sample fluctuations in GST get amplified, causing poor accuracy.

We would like both preparation and measurement fiducials to be \noemph{uniformly IC}, meaning that they span $\mathcal{B}(H)$ as uniformly as possible, and the smallest 
singular value of the Gram matrix is as large as possible.  There exists a single-qubit uniformly IC set with only 4 elements (the SIC-POVM), but it cannot be generated with Clifford operations and stabilizer states.  The smallest convenient uniformly IC set is the 6-element set of stabilizer states (the eigenstates of $X$, $Y$, and $Z$).  We choose six fiducial sequences so that, if the gates are ideal, they will prepare the stabilizer states exactly.  They are
\begin{equation}
\emptyset,\ G_x,\ G_y,\ G_xG_x,\ G_xG_xG_x,\ G_yG_yG_y,
\end{equation}
where $\emptyset$ indicates the null sequence (no gates).

Slightly imperfect gates will prepare states (and effects) that are close to the stabilizer states -- and therefore close to uniformly IC, and almost as effective in probing the operation of interest.  If the gates are sufficiently far from the targets, it can be detected by computing the singular values of the empirical Gram matrix, and then new fiducials can be chosen.  

Once the fiducials are defined, we need to define ``operations of interest'' for them to probe.  By sandwiching any such operation between an exhaustive set of 36 fiducial pairs, we are essentially doing process tomography on the operation (although the algorithm for incorporating these data into the GST fit is more complex than simple process tomography).

The obvious operations of interest are the gates themselves ($G_x$, $G_y$, and $G_i$).  By probing each gate tomographically, and repeating each sequence $N$ times, GST can estimate the gates to with $\pm \alpha/\sqrt{N})$ accuracy (for some constant $\alpha$).  To achieve higher accuracy, we do tomography on \noemph{powers} of the gates, by designating, \noemph{e.g.}, $G_x^{128}$ or $G_i^{32}$ as an operation of interest.  (Powers of 2 are chosen merely for convenience; any logarithmically spaced sequence of integer powers would work).

Repeating a gate $L$ times -- i.e., performing sequences of the form $F_j G_k^L F_i$ -- \noemph{amplifies} errors in the gate itself.  So, \noemph{e.g.}, if $G_x$ is actually a rotation by $\theta = \pi/2 + \epsilon$, then $G_x^{32}$ is a rotation by $32\epsilon$.  GST can now characterize \noemph{that} rotation to within $\pm \alpha/\sqrt{N}$, which equates to estimating $\theta$ to within $\alpha/(32\sqrt{N})$.  Raising gates to the $L$th power amplifies deviations by $L$, which in turn reduces estimation error by a factor of $L$.

However, simple repetition of $G_x$ does not amplify \noemph{every} error.  For example, suppose that $G_x$ is in fact a $\pi/2$ rotation, but around the wrong axis, corresponding to the unitary map
\begin{equation}
U = e^{-i(\pi/4)(\cos\epsilon X + \sin\epsilon Y)},
\end{equation}
as opposed to the target unitary $e^{-i(\pi/4)X}$ ($X$ and $Y$ indicate the Pauli operators $\sigma_x$ and $\sigma_y$).

This is a \noemph{tilt} error, and it is not amplified by $G_x^L$.  It's easy to see this by observing that $G_x^4 = \Id$, so the error cancels itself out after just 4 repetitions.  

More sophisticated sequences are needed to amplify tilt errors.  For this example, it is sufficient to probe $G_xG_y$.  Assuming (for now) that $G_y$ is a perfect $\pi/2$ rotation around $y$, $G_xG_y$ is a rotation by $2\pi/3 + \epsilon/\sqrt{3}$.  Therefore, performing $(G_xG_y)^L$ amplifies the deviation $\epsilon$ by a factor of $L$, and setting it as an operation of interest allows GST to estimate $\epsilon$ to within $\sqrt{3}\alpha/(L\sqrt{N})$.  The short sequence $G_xG_y$ is a \noemph{germ}, and repeating it $L$ times yields a \noemph{germ power sequence} that can be sandwiched between fiducials to equip GST with high sensitivity to the parameter $\epsilon$.

The general situation gets rapidly complicated -- \noemph{e.g.}, if $G_y$ is not perfect, then $G_xG_y$ alone cannot distinguish between $Y$ tilt in $G_x$ and $X$ tilt in $G_y$.  Each germ is sensitive to some nontrivial linear combination of gate set parameters.  To choose a set of germs, we list the possible germs (i.e., all reasonably short sequences), and for each germ $g$ we identify what linear combination of parameters it amplifies.  We do this by computing a Jacobian,
\begin{equation}\nabla_g^{(L)} \equiv \frac{1}{L}\left.\frac{\partial \left[\sigma(g)^L\right]}{\partial \vec{G}}\right|_{\vec{G}=\vec{G}_{\mathrm{target}}}, \label{eq:Jacobian}
\end{equation}
where $\sigma(g)$ is the \noemph{gate sequence product} for germ $g$ (obtained by just multiplying together the process matrices), and $\vec{G}$ is a vector containing all the parameters of the gate set (e.g., the elements of all the process matrices).

In the single-qubit case, $\sigma(g)$ is a $4\times4$ matrix, and $\vec{G}$ is 48-dimensional because it contains the elements of three $4\times4$ gate matrices.  Constraining all gates to be trace-preserving reduces the number of free parameters to 12 and 36 (respectively), so $\nabla_g$ is a $12\times36$ matrix.  Its 12 right singular vectors indicate linear combinations of gate set parameters that $\sigma(g)$ amplifies (when repeated $L$ times), and the corresponding singular values quantify how much they are amplified.  A zero singular value indicates a parameter that is not amplified at all (like the tilt error discussed above).  A \noemph{set} of germs $\{g_1\ldots g_N\}$ is, collectively, described by a Jacobian
\begin{equation}
J = \left(\begin{array}{c} \nabla_{g_1} \\ \nabla_{g_2} \\ \vdots \\ \nabla_{g_N} \end{array}\right).
\end{equation}

Our goal is to choose germs that provide high sensitivity at ``large'' values of $L$.  In practice, it is not useful to make $L$ larger than $1/\epsilon$, where $\epsilon$ is the rate of stochastic or depolarizing noise.  To select germs, however, we ignore this effect and make the simplifying assumption that the gates (and therefore $\sigma(g)$) are reversible (a good approximation when $\epsilon$ is small).
Under this assumption, it is possible to define the $L\to\infty$ limit of the Jacobian in Eq. \ref{eq:Jacobian}.  Using the product rule, and assuming that all the gates are unitary (and therefore $\sigma(g)^{-1} = \sigma(g)^\dagger$),
\begin{eqnarray}
\nabla_g^{(L)} &=& \frac{1}{L}\sum_{n=0}^{L-1} \sigma(g)^n \frac{\partial \sigma(g)}{\partial \vec{G}} \sigma(g)^{L-1-n} \\
&=& \left[ \frac{1}{L} \sum_{n=0}^{L-1} \sigma(g)^n \nabla_g^{(1)} (\sigma(g)^\dagger)^n \right]\sigma(g)^{-(L-1)}
\end{eqnarray}
As $L\to\infty$, the average over all powers $n$ of $\sigma(g)$ \noemph{twirls} $\nabla_g^{(1)}$.  By Schur's lemma, the effect of twirling is to project $\nabla_g^{(1)}$ onto the commutant of $\sigma(g)$ -- i.e., onto the subspace of matrices that commute with $\sigma(g)$.  Furthermore, multiplication by the unitary $\sigma(g)^{-(L-1)}$ is merely a change of basis, and has no effect on the right singular vectors or the singular values of $\nabla_g$.  So, up to an irrelevant change of basis:
\begin{equation}
\lim_{L\to\infty}{\nabla_g^{(L)}} = \Pi_{\sigma(g)}\left[ \nabla_g^{(1)} \right],
\end{equation}
where $\Pi_{\sigma(g)}$ is the projection onto the commutant of $\sigma(g)$.

This framework defines a notion of informational completeness for germs.  A set of germs $\{g_i\}$ is \noemph{amplificationally complete} (AC) if and only if the right singular rank of its Jacobian equals the total number of physically accessible (gauge-invariant) parameters in the gate set.  For a general set of 3 single-qubit trace-preserving gates, a gauge transformation is $G_k \to TG_kT^{-1}$ where $T$ is an invertible trace-preserving superoperator, so there are 12 gauge parameters, and 36-12=24 gauge-invariant parameters.  To build an AC set of germs, it is sufficient to add germs to the set until its Jacobian has rank 24.  By constructing a complete set of infinitesimal gauge transformations, we can actually construct the projector $\Pi_{g-i}$ onto the (local) space of gauge-invariant perturbations to the gate set $\vec{G}$.

We then optimize this set numerically, by adding and removing germs (taken from an exhaustive list of all short sequences), and only keeping a modification if it lowers a certain score function.  
(For single-qubit GST, we find it convenient to search over \noemph{all} germs of length $\leq6$.  However, this set of candidates need not be exhaustive.  (A larger gate set, for example, would generate a prohibitively large exhaustive candidate set.)  We have used randomly chosen subsets as candidate sets and gotten similar results.)
  The score function is
\begin{equation}
f(\{g_1\ldots g_k\}) = \frac{\Tr\left[(J^\dagger J)^{-1}\right]}{k}.
\end{equation}
This score estimates the mean squared error of estimation if a fixed number of counts are spread over the $k$ distinct germs.  Running this algorithm until it cannot improve the germ set any further produces the following set of 11 germs used in the final (March, 2015) GST runs (see Fig.~7): 

\begin{eqnarray}
&&G_x,\ G_y,\ G_i,\ G_xG_y,\nonumber\\
&&G_xG_yG_i,\ G_xG_iG_y,\ G_xG_iG_i, G_yG_iG_i,\nonumber\\
&&G_xG_xG_iG_y,\ G_xG_yG_yG_i,\ G_xG_xG_yG_xG_yG_y.
\end{eqnarray}

\begin{figure}[tbh!]
\includegraphics[width= 1.0 \linewidth]{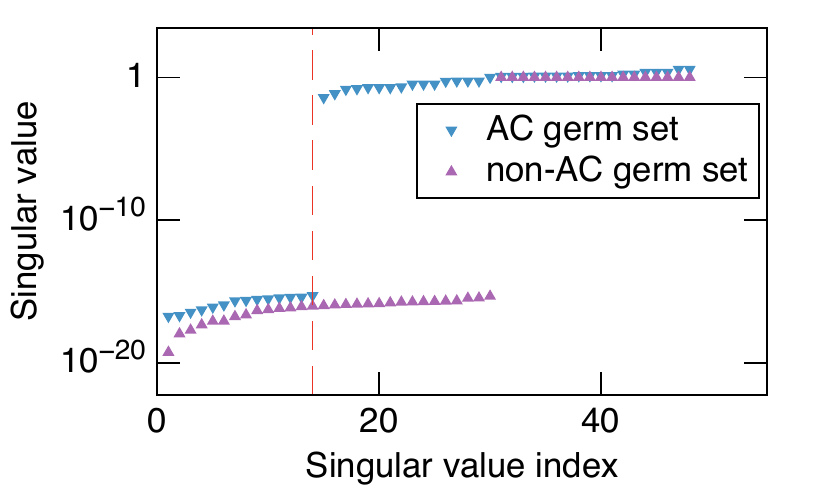}
\caption{\label{germs}
\textbf{Sensitivity analysis for germ selection.}
Here, we show the sorted singular values of the Jacobian matrix for different germ sets for the standard $\{G_X,G_Y,G_I\}$ gate set.  
Each singular value of the Jacobian corresponds to a gate set parameter; a large singular value indicates that the chosen germ set provides GST with sensitivity to that parameter.  Given that there are always experimentally inaccessible gauge parameters, it is impossible to be sensitive to all parameters.  
The dashed red line indicates the number of gauge parameters for this gate set (14).  
The blue triangles are singular values for a Jacobian with the amplificationally complete 11-germ set used for the March 2015 GST runs; all singular values corresponding to non-gauge parameters are large.  
The purple triangles are singular values for the 3-germ set containing just the bare gates $G_x$, $G_y$, and $G_i$.  
It is amplificationally incomplete, as indicated by the presence of near-zero singular values that correspond to non-gauge parameters.}
\end{figure}

\subsection{The GST gauge, and how to set it}

A gate set comprises: an initial density matrix $\rho$ (represented as a Hilbert-Schmidt vector), a measurement effect $E$ (represented as a Hilbert-Schmidt dual vector), and one or more gates $G_i$ (represented as superoperators).  But not every parameter in this representation is physically observable.  A gate set has intrinsic ``gauge'' degrees of freedom, because two distinct gate sets (or an entire manifold of them) can yield identical probabilities for \noemph{all} possible experiments.  \noemph{Gauge transformations} alter a gate set's elements without changing any observable probability.  They take the form
\begin{eqnarray}
\sbra{E} &\to& \sbra{E} M^{-1} \nonumber \\
\sket{\rho} &\to& M \sket{\rho} \nonumber \\
G_i &\to& M G_i M^{-1} \label{eq:GaugeTransform},
\end{eqnarray}
where $M$ is any invertible superoperator.  If (as usual) we consider only trace-preserving (TP) gate sets, then the corresponding necessary and sufficient condition for a gauge transformation to preserve this constraint is that $M$ be itself TP (i.e., its first row should be $(1,0,0,\ldots)$.

This gauge freedom makes it difficult to compare two gate sets, since two apparently-distinct gate sets may actually be equivalent.  Most of the metrics used to measure distance between two gates are not gauge-invariant (e.g.~fidelity, trace-norm distance, and diamond-norm distance are all gauge-variant).  So, while it would be ideal to work only with gauge-invariant metrics, we have very few metrics (and developing them and championing their adoption to the scientific community is beyond the scope of this work).  Instead, to generate meaningful metrics, we \noemph{gauge optimize} gate sets to make them as ``close'' as possible before computing metrics.

Given a gate set $\mathcal{G}$ and a target $\mathcal{G}'$, we transform $\mathcal{G}$ by $M$ (as above) where $M$ is chosen to optimize some criterion of ``closeness'' between $\mathcal{G}$ and $\mathcal{G}'$.  This is ``gauge optimization''.  The final output of GST is thus the gate set that is most similar to the target, according to some gauge-variant quantity, among a class of gauge-equivalent gate sets.  In the work reported here, we minimize (for convenience) a weighted Frobenius distance:
\begin{equation}
\begin{array}{l}
g(\mathcal{G},\mathcal{G}_0) = w_g \sum_i \lVert G_i  - G'_i \rVert^2 \\
 \quad + \, w_s \left( \lVert \rho - \rho' \rVert^2 
 + \lVert E - E' \rVert^2 \right),
\end{array}
\end{equation}
where $\lVert \cdot \rVert$ denotes the Frobenius norm, $G_i$ ranges over all gates in the set, and $w_g$ and $w_s$ are weighting factors.  The weight ratio $w_s/w_g$ allows us to fine-tune the relative contributions of discrepancies in logic gates and in SPAM.  This is important because their respective natural uncertainties are usually quite different; gates can be probed far more accurately than SPAM.  Thus, typically, $w_S/w_G \ll 1$; we weight the gate matrix elements more highly because they are known more precisely.  We use an iterative numerical method to find an $M$ that minimizes this quantity.

Each gate set is gauge-optimized as a whole; we report all metrics using gates in a single gauge.  It would be incorrect to separately optimize the gauge for different reported quantities (e.g. gauge-optimizing for the fidelity of a single gate and reporting each such best-fidelity separately).  Finally, we note that the process of gauge optimization against a reference gate set is sufficient to solve gauge ambiguity issues. That is, any quantity of interest that is not inherently gauge invariant (e.g., diamond norm) becomes so when this gauge optimization is performed. This numerical optimization process is not physically elegant, but is adequate for the practical applications we consider here.

\subsection{Error bars}
\label{sec:ErrorBars}
In interpreting the GST analyses (and in particular confirming the claim that we have demonstrated fault tolerance) it is necessary to assign error bars to gate set estimates (and derived quantities thereof).  For most GST-derived quantities, we use Hessian-based likelihood ratio (LR) confidence regions, while for RB-related quantities, we use parametric and non-parametric bootstrapping.  We also use parametric bootstrapped error bars as a sanity check on our Hessian-based LR confidence regions, and find them to be in good agreement.  Unless otherwise stated, all error bars indicate $\sim$$95\%$ ($2\sigma$) confidence intervals.

We employ two flavors of bootstrapping: parametric and non-parametric.  Both derive statistical quantities of interest from ensembles of simulated data sets, but these data sets are generated in different ways.
 
For the parametric bootstrap, ensembles of data sets are generated by first computing the GST estimate of the experimental data set in question, and then using this estimate to generate an ensemble of new data sets, each of which has the same experiments and number of shots per experiment as the actual experimental data set.

For the non-parametric bootstrap, ensembles of data sets are generated by simply resampling the experimental data set with replacement.  In both parametric and non-parametric bootstraps, we typically generate an ensemble of 100 data sets, to ensure good statistics.

GST is used to map each resampled dataset to a gate set estimate.  Each gate set is gauge-optimized to match the experimental GST estimate as closely as possible.  Then, from this ensemble of gauge-optimized gate sets, any statistical quantity of interest (such as standard deviation) may be calculated for process matrix elements or for derived quantities such as  diamond norm.

\begin{figure}[tb]
\includegraphics[width= 1.0 \linewidth]{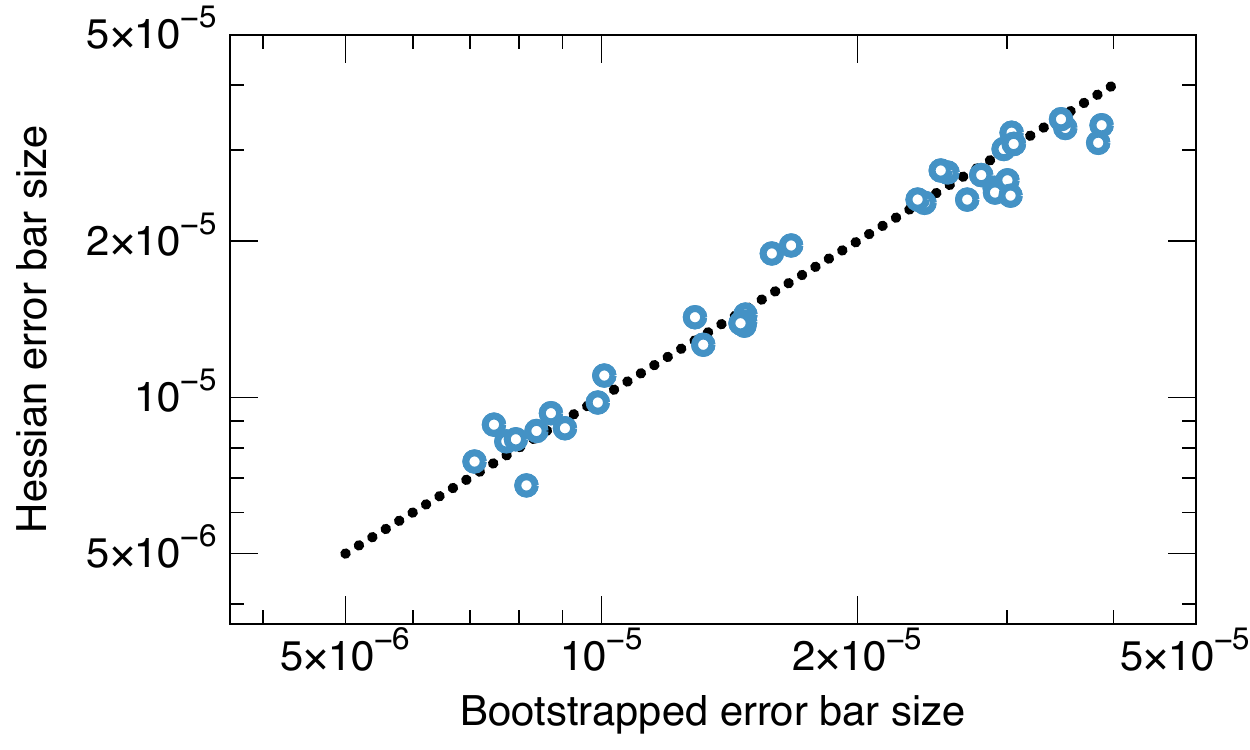}
\caption{\label{fig:ErrorBars}
\textbf{Comparison of error bar generation techniques.}
When computing error bars on GST estimates in this manuscript, we typically use likelihood ratio (LR) confidence regions, computed using the Hessian of the log likelihood function.
However, another common approach is through parametric bootstrapping. 
Here, we show a log-log scatter plot of error bars on gate elements from GST estimate of our data from 30 March 2015.  The x-axis corresponds to error bars calculated via parametric bootstrapping, whereas the y-axis corresponds to likelihood ratio (LR) confidence regions computed using the Hessian of the log likelihood function.  The dotted line corresponds to $y=x$.  Both methods are described in the methods section ``Error Bars''.  The strong correlation shown here demonstrates the consistency between parametric bootstrapping techniques and LR confidence regions.}
\end{figure}

We use bootstrapped error bars for two purposes.  First, they serve as a sanity check on the more rigorous (but tricky to implement) LR confidence regions described in the following subsection.  In Figure~\ref{fig:ErrorBars} we compare the performance of parametric bootstrapping to the LR method, and see good agreement.  Second, we use bootstrapping to put error bars on experimental RB decay rates.  These are model-free and therefore not amenable to LR confidence regions.  Error bars on experimental RB decay rate error bars were calculated via non-parametric bootstrapping, while error bars on simulated RB decay rates were calculated via parametric bootstrapping on the underlying GST estimate used to generate the RB data.  

Bootstrapping is a very general method for generating error bars, but it is (1) not always reliable, (2) subject to small-sample errors unless very many Monte Carlo samples are generated, and (3) quite time-consuming (up to 24 hours of computer time were required to generate the 100 samples used for this paper).  \noemph{Likelihood ratio (LR) confidence regions} \cite{RBK12} are preferable in most ways, and we use them as our primary source for ``error bars''.

The basic theory for LR confidence regions, as applied to quantum tomography, can be found in Ref. \cite{RBK12}.  Confidence regions have a solid (if often misunderstood) statistical meaning:  if an estimator generates confidence regions with a confidence level of $1-\alpha$, then with probability at least $1-\alpha$ (taken over the ensemble of all possible datasets), the confidence region assigned by the estimator will contain the true parameter value.  This does \noemph{not} mean ``Given \noemph{particular} error bars, the probability that they contain the truth is $1-\alpha$,'' as there is no random variable to take a probability with respect to once the estimate has been assigned.

As implemented here, GST has two convenient properties.  First, it yields a likelihood function that is well-approximated by a Gaussian (because the total number of samples is quite large).  Second, it involves no explicit constraints, meaning that the MLE is never squashed against a boundary (as it often is in standard state and process tomography, where the positivity constraint is critical to ensuring a physically valid estimate).  These properties mean that we can approximate the loglikelihood function by a quadratic function, whose shape is given by the Hessian (matrix of 2nd derivatives) of $\log\mathcal{L}$ at the MLE.  This Hessian defines a covariance tensor in gate set space, which (when scaled by an appropriate factor) defines an ellipsoid that is a valid $1-\alpha$ confidence region.

Writing down this ellipsoid explicitly (as a covariance tensor) is possible, but not useful in practice.  Instead, we use it to define error bars (confidence intervals) for all relevant scalar quantities (including fidelities, diamond norms, gate matrix elements, etc).

Let $f(\mathcal{G})$ be a scalar function of a gate set.  We define a 95\% confidence interval around the best-estimate value of $f^* = f(\mathcal{G}_{\mathrm{best}})$ by computing
\begin{equation}
\delta f = \sqrt{ (\nabla f)^{\dagger} \cdot (P(H)/C_1)^{-1} \cdot \nabla f  } \label{eq:HessianErrorBars}
\end{equation}
where $P(H)$ is the Hessian projected onto the (local) space of non-gauge gate set parameters, and we have linearized $f(\mathcal{G}) \approx f_0 + \nabla f \cdot (\mathcal{G}-\mathcal{G}_{\mathrm{best}})$.  $C_1$ is a scalar constant which satisfies $\mathrm{CDF}_1(C_1) = 0.95$, where $\mathrm{CDF}_1$ is the cumulative density function of the $\chi^2_1$ probability distribution.  With $\delta f$ so defined, $f^* \pm \delta f$ specifies the 95\% confidence interval for $f$.  Within the linear approximation to $f$, which is valid for small $\delta f$, this interval corresponds to minimizing and maximizing the value of $f$ over the contour of the log-likelihood corresponding to a 95\% confidence interval \noemph{if} the log-likelihood had a single parameter.  

We emphasize that this does not construct a 95\% confidence \noemph{region}.  There are roughly 34 gauge-invariant parameters in a gate set; the threshold used here implies 95\% confidence \noemph{intervals} for each of them.  The resulting region contains the truth only if every one of the intervals contains its parameter, which occurs with probability at least $0.95^{34} \approx 17\%$.  

We believe this is a more meaningful way to report ``error bars'' than to report a 95\% confidence region for the entire gate set.  For one thing, it is consistent with the error bars reported by the bootstrap (which yields standard errors for each parameter independently, and would have to be expanded significantly to represent a joint confidence region).  Empirically, we find that definition \ref{eq:HessianErrorBars} correlates closely with the $2\sigma$ error bars on gate elements computed by parametric bootstrapping (see Fig.~\ref{fig:ErrorBars}).  Furthermore, we use the confidence region primarily to report uncertainties on single quantities (e.g. diamond norms), independent of the others.

\subsection{Data availability}
The GST and RB analysis in this paper was performed using the open-source software pyGSTi (python Gate Set Tomography implementation) \cite{pygsti}, which was developed for this work.  All datasets and analysis scripts necessary to reproduce the results presented here are available online as supplemental information at \url{https://doi.org/10.5281/zenodo.231329}.

\bibliographystyle{naturemag}
\bibliography{citations}

\section{Acknowledgements}

Sandia National Laboratories is a multi-program laboratory managed and operated by Sandia Corporation, a wholly owned subsidiary of Lockheed Martin Corporation, for the U.S. 
Department of Energy's National Nuclear Security Administration under contract DE-AC04-94AL85000.  The authors thank Travis Scholten and Jonathan Gross for assistance with data visualization and Kevin Young for providing code for diamond norm computation.  JKG gratefully acknowledges support from the Sandia National Laboratories 
Truman Fellowship Program, which is funded by the Laboratory Directed Research and Development (LDRD) program.  This research was funded, in part, by the Office of the Director of 
National Intelligence (ODNI), Intelligence Advanced Research Projects Activity (IARPA). All statements of fact, opinion or conclusions contained herein are those of the authors and 
should not be construed as representing the official views or policies of IARPA, the ODNI, or the U.S. Government.

\section{Author contributions}
RBK, JKG, EN, and KR contributed to the theoretical development of GST and the code implementation.  JM, KF, and PM performed the experiments. All authors discussed the results and wrote the manuscript.

\section{Competing interests}
The authors declare no competing financial interests.

\end{document}